\documentclass[traditabstract]{aa}
\usepackage{natbib, txfonts,graphicx,amssymb,url}

\begin{document}
\title{Wide-band Simultaneous Observations of Pulsars: Disentangling Dispersion Measure and Profile Variations}
\author{T.~E.~Hassall\inst{\ref{jod}}\and
B.~W.~Stappers\inst{\ref{jod}} \and
J.~W.~T.~Hessels\inst{\ref{astron} \and \ref{uva}}\and
M. ~Kramer\inst{\ref{mpifr} \and \ref{jod}}\and
A. ~Alexov\inst{\ref{uva}}\and
K.~Anderson\inst{\ref{uva}}\and
T.~Coenen\inst{\ref{uva}} \and
A.~Karastergiou\inst{\ref{ox}}\and
E.~F.~Keane\inst{\ref{mpifr}}\and
V. ~I.~Kondratiev\inst{\ref{astron}}\and
K.~Lazaridis\inst{\ref{mpifr}}\and
J.~van Leeuwen\inst{\ref{astron} \and \ref{uva}} \and
A.~Noutsos\inst{\ref{mpifr}}\and
M.~Serylak\inst{\ref{nancay} \and \ref{cnrs}}
C.~Sobey\inst{\ref{mpifr}}\and
J.~P.~W.~Verbiest\inst{\ref{mpifr}}\and
P.~Weltevrede\inst{\ref{jod}}\and
K.~Zagkouris\inst{\ref{ox}}\and
R.~Fender\inst{\ref{soton}}\and
R.~A.~M.~J.~Wijers\inst{\ref{uva}}\and
L.~B\"ahren\inst{\ref{uva}}\and
M.~E.~Bell\inst{\ref{soton} \and \ref{sydney}} \and 
J.~W.~Broderick\inst{\ref{soton}} \and
S.~Corbel\inst{\ref{paris} \and \ref{IUF}}
E.~J.~Daw\inst{\ref{sheff}}\and
V.~S.~Dhillon\inst{\ref{sheff}}\and
J.~Eisl\"offel\inst{\ref{tls}} \and 
H.~Falcke\inst{\ref{nijmegen} \and \ref{astron} \and \ref{mpifr}} \and 
J.-M.~Grie{\ss}meier\inst{\ref{cnrs}} \and 
P.~Jonker\inst{\ref{sron} \and \ref{nijmegen}\and \ref{harv}}
C.~Law\inst{\ref{berkley} \and \ref{uva}} \and           
S.~Markoff\inst{\ref{uva}} \and
J.~C.~A.~Miller-Jones\inst{\ref{curtin} \and \ref{uva}} \and 
R.~Osten\inst{\ref{stsi}}\and
E.~Rol\inst{\ref{uva}}\and
A.~M.~M.~Scaife\inst{\ref{soton}}\and
B.~Scheers\inst{\ref{uva} \and \ref{cwi}} \and 
P.~Schellart\inst{\ref{nijmegen}}\and 
H.~Spreeuw\inst{\ref{uva}} \and 
J.~Swinbank\inst{\ref{uva}} \and 
S.~ter Veen\inst{\ref{nijmegen}}\and
M.~W.~Wise\inst{\ref{astron} \and \ref{uva}} \and 
R.~Wijnands\inst{\ref{uva}}
O.~Wucknitz\inst{\ref{ubonn}} \and 
P.~Zarka\inst{\ref{meudon}} \and 
A.~Asgekar\inst{\ref{astron}} \and 
M.~R.~Bell\inst{\ref{mpifa}} \and
M.~J.~Bentum\inst{\ref{astron}} \and 
G.~Bernardi\inst{\ref{harv}}\and
P.~Best\inst{\ref{roe}} \and 
A.~Bonafede\inst{\ref{bremen}}\and
A.~J.~Boonstra\inst{\ref{astron}}\and
M.~Brentjens\inst{\ref{astron}} \and 
W.~N.~Brouw\inst{\ref{kapteyn}} \and 
M.~Br\"uggen\inst{\ref{bremen}} \and 
H.~R.~Butcher\inst{\ref{astron} \and \ref{anu}} \and      
B.~Ciardi\inst{\ref{mpifa}} \and 
M.~A.~Garrett\inst{\ref{astron} \and \ref{leiden}} \and 
M.~Gerbers\inst{\ref{astron}} \and 
A.~W.~Gunst\inst{\ref{astron}} \and 
M.~P.~van Haarlem\inst{\ref{astron}} \and 
G.~Heald\inst{\ref{astron}} \and 
M.~Hoeft\inst{\ref{tls}} \and 
H.~Holties\inst{\ref{astron}} \and 
A.~de Jong\inst{\ref{astron}}\and
L.~V.~E.~Koopmans\inst{\ref{kapteyn}} \and 
M.~Kuniyoshi\inst{\ref{mpifr}}\and
G.~Kuper\inst{\ref{astron}} \and 
G.~M.~Loose\inst{\ref{astron}} \and 
P.~Maat\inst{\ref{astron}} \and 
J.~Masters\inst{\ref{nrao}}\and
J.~P.~McKean\inst{\ref{astron}} \and 
H.~Meulman\inst{\ref{astron}}\and
M.~Mevius\inst{\ref{astron}}\and
H.~Munk\inst{\ref{astron}}\and
J.~E.~Noordam\inst{\ref{astron}} \and 
E.~Orr\'u\inst{\ref{nijmegen}}\and
H.~Paas\inst{\ref{astron}}\and
M.~Pandey-Pommier\inst{\ref{lyon}} \and 
V.~N.~Pandey\inst{\ref{astron}}\and
R.~Pizzo\inst{\ref{astron}}\and
A.~Polatidis\inst{\ref{astron}} \and 
W.~Reich\inst{\ref{mpifr}} \and 
H.~R\"ottgering\inst{\ref{leiden}} \and 
J.~Sluman\inst{\ref{astron}} \and 
M.~Steinmetz\inst{\ref{aip}} \and 
C.~G.~M.~Sterks\inst{\ref{groningen}} \and  
M.~Tagger\inst{\ref{cnrs}} \and 
Y.~Tang\inst{\ref{astron}} \and 
C.~Tasse\inst{\ref{meudon}}\and
R.~Vermeulen\inst{\ref{astron}} \and 
R.~J.~van Weeren\inst{\ref{leiden} \and \ref{astron}} \and
S.~J.~Wijnholds\inst{\ref{astron}} \and 
S.~Yatawatta\inst{\ref{kapteyn}}
}

\institute{Jodrell Bank Centre for Astrophysics,
School of Physics and Astronomy,
The University of Manchester,
Manchester M13 9PL, UK\label{jod}
\email{tomehassall@gmail.com}
\and ASTRON, the Netherlands Institute for Radio Astronomy, Postbus 2, 7990 AA Dwingeloo, The Netherlands\label{astron}
\and Astronomical Institute ``Anton Pannekoek'', University of Amsterdam, Science Park 904, 1098 XH Amsterdam, The Netherlands\label{uva}
\and Max-Planck-Institut f\"ur Radioastronomie, Auf dem H\"ugel 69, 53121 Bonn, Germany\label{mpifr}
\and Astrophysics, University of Oxford, Denys Wilkinson Building, Keble Road, Oxford OX1 3RH\label{ox}
\and Station de Radioastronomie de Nan\c cay, Observatoire de Paris, CNRS/INSU, 18330 Nan\c cay, France\label{nancay}
\and Laboratoire de Physique et Chimie de l'Environnement et de l'Espace, LPC2E UMR~7328 CNRS, 45071 Orl\'eans Cedex 02, France\label{cnrs}
\and School of Physics and Astronomy, University of Southampton, Southampton, SO17 1BJ, UK\label{soton}
\and Sydney Institute for Astronomy, School of Physics, The University of Sydney, NSW 2006, Australia\label{sydney}
\and Universit\'{e} Paris 7 Denis Diderot and Service d'Astrophysique, UMR AIM, CEA Saclay, F-91191 Gif sur Yvette, France \label{paris}
\and Institut Universitaire de France, 75005 Paris, France \label{IUF}
\and Department of Physics \& Astronomy, University of Sheffield, Sheffield S3 7RH, United Kingdom\label{sheff}
\and Th\"uringer Landessternwarte, Sternwarte 5, D-07778 Tautenburg, Germany\label{tls}
\and Department of Astrophysics, IMAPP, Radboud University Nijmegen, P.O. Box 9010, 6500 GL Nijmegen, The Netherlands\label{nijmegen}
\and SRON, Netherlands Institute for Space Research, Sorbonnelaan 2, 3584 CA, Utrecht, The Netherlands\label{sron}
\and Harvard-Smithsonian Center for Astrophysics, Garden Street 60, Cambridge, MA 02138, USA.\label{harv}
\and Radio Astronomy Lab, UC Berkeley, CA, USA\label{berkley}
\and International Centre for Radio Astronomy Research - Curtin University, GPO Box U1987, Perth, WA 6845, Australia\label{curtin}
\and Space Telescope Science Institute, Baltimore, MD 21218, USA\label{stsi}
\and Centrum Wiskunde \& Informatica (CWI), PO Box 94079, 1090 GB, Amsterdam, the Netherlands \label{cwi}
\and Argelander-Institut f\"ur Astronomie, Universit\"at Bonn, Auf dem H\"ugel 71, 53121, Bonn, Germany\label{ubonn}
\and LESIA, Observatoire de Paris, CNRS, UPMC Universit\'{e} Paris-Diderot, 5 place Jules Janssen, 92195 Meudon, France\label{meudon}
\and Max Planck Institute for Astrophysics, Karl Schwarzschild Str. 1, 85741 Garching, Germany\label{mpifa}
\and Institute for Astronomy, University of Edinburgh, Royal Observatory of Edinburgh, Blackford Hill, Edinburgh EH9 3HJ, UK\label{roe}
\and Jacobs University Bremen, Campus Ring 1, 28759 Bremen, Germany\label{bremen}
\and Kapteyn Astronomical Institute, PO Box 800, 9700 AV Groningen, The Netherlands\label{kapteyn}
\and Mt Stromlo Observatory, Research School of Astronomy and Astrophysics, Australian National University, Weston, A.C.T. 2611, Australia\label{anu}
\and Leiden Observatory, Leiden University, PO Box 9513, 2300 RA Leiden, The Netherlands\label{leiden}
\and NRAO Headquarters, 520 Edgemont Road, Charlottesville, VA 22903, USA\label{nrao}
\and Centre de Recherche Astrophysique de Lyon, Observatoire de Lyon, 9 av Charles Andr\'e, 69561 Saint Genis Laval Cedex, France\label{lyon}
\and Leibniz-Institut fŸr Astrophysik Potsdam (AIP), An der Sternwarte 16, 14482 Potsdam, Germany\label{aip}
\and Center for Information Technology (CIT), University of Groningen, The Netherlands\label{groningen}
}

\titlerunning{Wide-band Simultaneous Observations of Pulsars}

\abstract{Dispersion in the interstellar medium is a well known phenomenon that follows a simple relationship, which has been used to predict the time delay of dispersed radio pulses since the late 1960s. We performed wide-band simultaneous observations of four pulsars with LOFAR (at 40--190~MHz), the 76-m Lovell Telescope (at 1400~MHz) and the Effelsberg 100-m Telescope (at 8000~MHz) to test the accuracy of the dispersion law over a broad frequency range. In this paper we present the results of these observations  which show that the dispersion law is accurate to better than 1 part in $10^5$ across our observing band. We use this fact to constrain some of the properties of the ISM along the line-of-sight and use the lack of any aberration or retardation effects to determine upper limits on emission heights in the pulsar magnetosphere. We also discuss the effect of pulse profile evolution on our observations, and the implications that it could have for precision pulsar timing projects such as the detection of gravitational waves with pulsar timing arrays.}


\maketitle

\section{Introduction}
\label{sec:intro}
As radio emission passes through the interstellar medium (ISM) it interacts with electrons, which cause it to be dispersed. This is observable through pulsed emission, where a low-frequency pulse is delayed with respect to the same pulse at higher frequencies. 
For a pulse at frequency $\nu$ (frequency will be given in MHz here and throughout the paper) travelling a distance, $D$, through an unmagnetised ionised gas, the dispersive delay with respect to infinite frequency, $\Delta t_{\mathrm{DM}}$, is given by:
\begin{equation}
\Delta t_{\mathrm{DM}} = \left( \int_0^D \frac{\mathrm{d}l}{v_g} \right) -\frac{D}{c} ,
\label{eq:disp}
\end{equation}
where $v_g$ is the group velocity:
\begin{equation}
v_g = c\sqrt{1-\left(\frac{\nu_p}{\nu}\right)^2} ,
\end{equation} 
$\nu_p$ is the plasma frequency:
\begin{eqnarray}
\nu_p &=& \sqrt{\frac{e^2n_e}{\pi m_e}} ~~\mathrm{(in~cgs~units)} \\
	  &\approx& 8.98\times10^{-3} \sqrt{n_e} ~\mathrm{MHz}~, 
\end{eqnarray}
$c$ is the speed of light, $m_e$ and $e$ are the charge and mass of an electron respectively, and $n_e$  is the electron density in cm$^{-3}$. Normally, this time delay can be approximated using the first term in the Taylor expansion of $1/v_g$, giving the cold dispersion law\footnote{Note that often, the more precise value of $2.410332\times10^{-4}$ is used as the constant in Equation~\ref{eq:disp1}, however in this paper we follow the convention of using $2.41\times10^{-4}$. This is only important when considering absolute values of DM. It should also be noted that this approximation only holds when the observing frequency is much greater than both the local plasma frequency and electron gyrofrequency.}:
\begin{equation}
\label{eq:disp1}
\Delta t_{\mathrm{DM}} = \frac{\mathrm{DM}}{2.41\times10^{-4} \nu^2}~\mathrm{s}  ,
\end{equation}
where DM is the dispersion measure, the integrated column density of free electrons in pc~cm$^{-3}$. This relation was used as early as 1968 to accurately predict the dispersive time delay to within 1 part in 3000 between 40~MHz and 430~MHz \citep{tzd68}. However, as ever higher timing precision is required for projects like using pulsars for gravitational wave detection \citep{jhlm05} it is important that any second-order effects of the ISM, such as refractive delays, DM variations and delays associated with pulse broadening from scattering \citep{fc90, cs10}, are also studied and understood fully \citep{yhc+08,hs08}. Most of these proposed effects have strong frequency dependencies, with scaling indices between $\nu^{-3}$ and $\nu^{-4}$, and are therefore most prominent at low frequencies.

As well as second-order ISM effects, there are other proposed frequency-dependent effects, such as propagation from within the pulsar magnetosphere \citep{mic91}, super-dispersion \citep{sm85,kuz86,smi88,klll08} and aberration and retardation \citep{cor78}. The ionosphere also contributes slightly to the DM but typically the total number of electrons along a path through the ionosphere is less than 100 TEC units \citep[e.g.][]{lwnz09}, which corresponds to a DM of just $3.24\times10^{-5}$~pc~cm$^{-3}$. In practice, this contribution is small and follows a $\nu^{-2}$ law, making it indistinguishable from the dispersive delay of the ISM.

As the sizes of these effects have not yet been determined, their frequency dependence could introduce systematic errors when data from different frequencies are combined. The size of these effects could also potentially vary with time. For example, if the distribution of free electrons along the line-of-sight changes, the magnitude of some of these effects will vary, introducing further errors into pulsar timing data. 

Another potential source of error arises from pulse profile evolution. Most (if not all) pulsars show variation in the shape of their pulse profile as a function of frequency \citep{cra70}. This change can be anything from a slight broadening of the pulse, or difference in the relative positions of the components, to components appearing or disappearing completely.

Even though wide-band simultaneous observations have been possible to perform for a long time \citep[see for example][]{pw92,kis+98,hr10} the relatively narrow frequency bands compared to the separation interval between frequencies have meant identifying exactly how the components evolve has been difficult, particularly at the lowest frequencies. One of the reasons for this is that profile evolution, which only manifests itself clearly over very wide frequency ranges, is difficult to disentangle from dispersion (dispersion is normally removed by fitting the data with a $\nu^{-2}$ power law which mimics Equation \ref{eq:disp1}). For example, in \cite{kis+98} the authors suggest that PSR B0809+74 appears to have an extra, non-dispersive delay of $\sim30$~ms at 10~GHz, though this could alternatively be explained by pulse profile evolution.

To calculate pulse times of arrival (TOAs), the data are cross-correlated with a `template', a smoothed model of the pulse profile. The peak of this cross-correlation spectrum is used to determine when the pulse arrives relative to the template, and the phase of the Fourier transform of the cross-correlation is then used to measure the phase offset. If the pulse profile evolves significantly with frequency the shape of the data and the template can be slightly different. These subtle changes in shape can introduce frequency-dependent errors into TOAs if they are not accounted for properly. \cite{amg07} simulated the effects of pulse profile evolution on measurements of DM. They found that using a template which is slightly different from the shape of the data can cause a gradient in the phase of the cross-correlation, which causes an offset in the TOA. 

Some of the simulated errors, in the work of Ahuja et al, due to this effect in normal (slow) pulsars were as large as milliseconds, although the magnitude of the effect  on real data has not been studied in detail until the work presented here. The size of this effect should scale linearly with pulse period, and so it is expected to be much smaller for millisecond pulsars, which is why sub-microsecond timing precision is already common \citep[for example, see][]{vbc+09}. But, with the introduction of the first ultra-broadband receivers to be used for pulsar timing \citep[e.g.][]{drd+08, jwm+10}, which will be capable of observing with large fractional bandwidths even at frequencies of a few gigahertz, pulse shape variation across the band will become far more pronounced, and this effect will become more apparent.

We took observations of four bright pulsars (pulsars B0329+54, B0809+74, B1133+16 and B1919+21) \emph{simultaneously} at multiple frequencies with the LOw Frequency ARray (LOFAR, see van Haarlem et al. in prep.), the 76-m Lovell Telescope and the Effelsberg 100-m Telescope, spanning a frequency range between 40 and 8350~MHz to try to constrain the properties of the ISM and test the cold plasma dispersion law. Details of these observations are given in Section \ref{sec:obs}. Our observations cover a very large frequency range simultaneously and each recorded band has a significant fractional bandwidth, also making them useful for studying pulse profile evolution. LOFAR provides the largest fractional bandwidth which has ever been possible at the lowest radio frequencies observable from Earth. 

The observations from both LOFAR observing bands were also taken truly simultaneously, both starting and ending at exactly the same time\footnote{There are small clock offsets between the LOFAR stations at this stage in construction, but these are less than 20~ns.}, and passing through the same hardware so no timing model was required to align the profiles precisely (as opposed to using overlapping observations with a timing model). This part of our frequency range provides the best test of steeply scaling frequency-dependent delays and this allows us to place strong constraints on the magnitude of the second-order ISM effects.  Any deviation from the $\nu^{-2}$ law larger than a few milliseconds would be clearly visible over this large bandwidth and at these low frequencies. 

Similar observations have been carried out previously by \cite{tzd68}, who were able to set a surprisingly good limit that the error in the dispersion equation, $\Delta t < 3\times10^{-4}$~s between 40~MHz and 430~MHz just six months after the discovery of pulsars in 1967. Since then similar work has been done by \cite{sm85}, who found evidence for excess dispersion at low frequencies. This was also supported by \cite{him+91}, who noticed a difference in the DM values determined from aligning single pulses and those determined from aligning the average pulse profile.  These delays at low frequencies have been termed `super-dispersion' and attributed to magnetic sweepback in the pulsar magnetosphere \citep{shi83}. Subsequent work by \cite{pw92}  and \cite{agmk05} found contradictory results, with no evidence for extra dispersion at low frequencies, but some pulsars (B0525+21, B1642-03 and B1237+25) showed an increased dispersion measure at high frequencies. These high frequency delays were explained by propagation delays in the pulsar magnetosphere.

In this paper, after detailing the observations (Section~\ref{sec:obs}), our analysis (Section~\ref{sec:analysis}), and the simulations used to determine upper limits on the magnitude of the delay present in our data (Section~\ref{sec:sims}); we determine the effect of the ISM on pulsar timing and extract information about the composition of the ISM (Section~\ref{sec:ism}) and the pulsar magnetosphere (Section~\ref{sec:mag}). We also use these data to determine how much of an impact profile evolution has on pulsar timing and use the pulse profiles in each of the frequency bands to construct a model of how the pulse profile evolves with frequency (Section~\ref{sec:prof}). Finally, we summarise our main findings in Section~\ref{sec:disc}.

\section{Observations}
\label{sec:obs}
On 11 December 2009 we observed four pulsars using LOFAR, the 76-m Lovell Telescope and the Effelsberg 100-m Telescope simultaneously. LOFAR is an international interferometric telescope, comprised of many thousands of dipole antennas grouped into `stations' and operating in the lowest four octaves of the `radio window' visible from the Earth's surface. The LOFAR stations are arranged in a sparse array, spread across Europe, with a dense core region located in the Netherlands. LOFAR operates in two frequency bands which straddle the FM radio band.  The Low Band Antennas (LBAs)  can record 48~MHz of bandwidth between 10 and 90~MHz and the High Band Antennas (HBAs) can record 48~MHz of bandwidth between 110 and 240~MHz.  For a full description of how LOFAR is used for pulsar observations see \cite{sha+11} and for a general LOFAR description see van Haarlem et al. (in prep.).

At the time of the observations, LOFAR was still under construction so only single stations were used to record data and the LBAs were only able to record 36~MHz of bandwidth. A core station (CS302, which consists of 48 HBA tiles) was used to observe in the LOFAR high band between 138.9709 and 187.4084~MHz and an international station (DE601, based at the Effelsberg Radio Observatory and consisting of 96 LBAs) was used to observe between 42.0959 and 78.4119~MHz in the low band. It is worth noting that the LOFAR stations which were used were not yet internally calibrated and hence the sensitivity of the observations presented here is significantly lower  compared to what is now possible.

Data were taken simultaneously with the 76-m Lovell Telescope, using the digital filterbank backend in search mode, at a central frequency of 1524~MHz with 512~MHz bandwidth \citep[see][for details]{hlk+04}. Simultaneous data were also taken with the Effelsberg 100-m Telescope, using the Effelsberg-Berkeley Pulsar Processor to coherently dedisperse the data online, at a central frequency of 8350~MHz with 1000~MHz bandwidth \citep[see][for details]{bdz+97}. The observational parameters are summarised in Tables \ref{tab:observations} and \ref{tab:psrobs}. 

\begin{table}
\caption{Data characteristics. Shown are the centre frequency ($\nu$), bandwidth ($B$),  number of channels ($N_\mathrm{chan}$), and the sampling time ($T_\mathrm{samp}$) for the data of each of the four telescopes used in the simultaneous observations.}
\centering
\label{tab:observations}
\begin{tabular}{ccccc}
\hline	
	 	    & $\nu$~(MHz) & $B$~(MHz)         & $N_\mathrm{chan}$ & $T_\mathrm{samp}$~(ms) \\ \hline
	DE601 &   60.15625    & 36.328125   & 2976 		& 1.31072\\
	CS302 & 163.28125    & 48.4375 	 & 3968 		& 1.31072 \\
	Lovell   & 1524.0 	    & 512 		 & 512   		& 1.0\\
	Effelsberg	& 8350.0	    & 1000	 	& 1$^a$		& $\sim1.0^b$\\ \hline
\end{tabular}
\tablefoot{
\tablefoottext{a}{Data was written as a single dedispersed timeseries.}
\tablefoottext{b}{This varied depending on the source. The pulse period was divided into 1024 bins.}
}
\end{table}

\begin{table}
\centering
\caption{Pulsar characteristics. Given are the integration time ($T_\mathrm{int}$), catalogue value for dispersion measure \citep[DM,][]{hlk+04},  the pulse period ($P$, determined from regular timing observations from the Lovell telescope at Jodrell Bank Observatory), the LOFAR observation ID and the number of bins across the pulse profile ($N_\mathrm{bin}$ for each of the pulsars observed.)}
\begin{tabular}{cccccc}
	\hline	
	 	           &$T_\mathrm{int}$& DM     & $P$   & LOFAR    	&  $N_\mathrm{bin}^a$\\ 
		           &(s) 	 & (pc cm$^{-3}$)     & (s)   &   Obs ID 	&		\\ \hline
	B0329+54 &   7200      & 26.833	 & 0.714536	  & L2009\_16116 	& 256\\
	B0809+74 &   5400  	    & 6.116 	 & 1.292209 	 &  L2009\_16102	& 512\\
	B1133+16 &   10800	    & 4.864		 & 1.187799   	 &  L2009\_16100	& 512\\
	B1919+21 &   5400      & 12.455	 & 1.337360	 &  L2009\_16104	& 512\\ \hline
\end{tabular}
\tablefoot{
\tablefoottext{a}{Although the pulsars in our sample were observed with higher time resolution, the data were downsampled so that each observation had the same number of bins in the pulse profile, so that the data from different frequencies could be compared directly.}
}
\label{tab:psrobs}
\end{table}

\begin{figure*}
\centering
\includegraphics[width=\linewidth, trim = 1cm 2cm 1cm 8cm, clip=true]{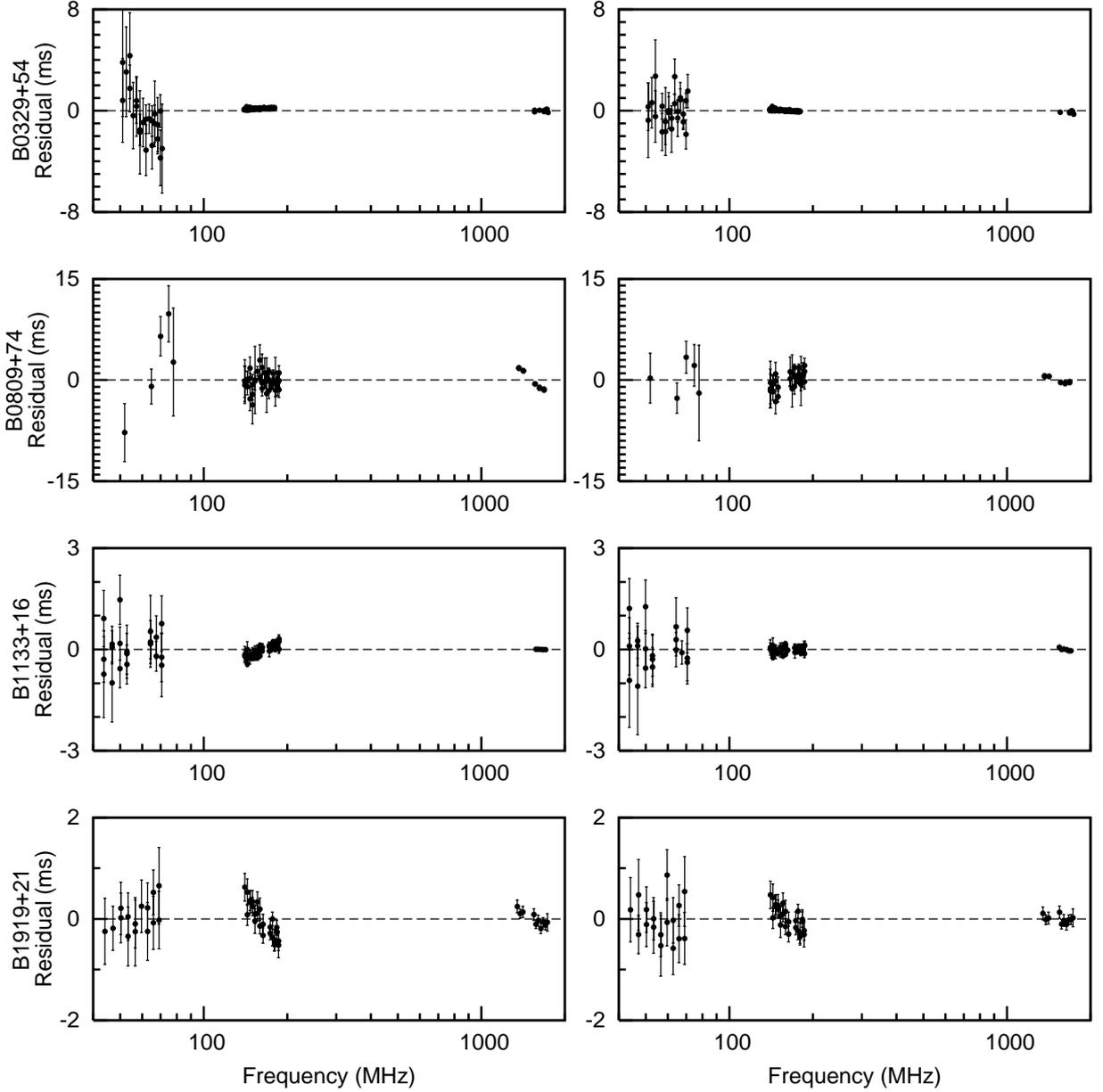}
\caption{Comparison of timing residuals (TOAs subtracted from a model of the pulsar's expected TOAs) obtained using a single template for each frequency band (left), and using a frequency-dependent template (right), plotted against $\nu$ on a logarithmic scale. The residuals from the static templates show significant deviations from white noise. These systematic errors are mostly removed with the frequency-dependent templates and the residuals appear straighter, and agree on a single value of DM (apart from those of PSR B1133+16, see text for details). `Gaps' in the frequency coverage are from subbands which have been removed because they contain strong RFI. The pulsars used are (from top to bottom) B0329+54, B0809+74, B1133+16 and B1919+21.}
\label{fig:resid}
\end{figure*}

In this paper, we concentrate mainly on the total intensity profiles of pulsars, and only Stokes I data were recorded with each instrument (LOFAR, the Lovell telescope and the Effelsberg telescope). However, if there is a significant gain difference between the two hands of (either linear or circular) polarisation, or leakage, and the pulsar is strongly polarised then there may be distortions in the profile. Instrumental polarisation from the Lovell and Effelsberg telescopes is known to be small \citep[see e.g.][]{mmv+08, gl98}, but at the time of these observations, the LOFAR polarisation data was uncalibrated. We note however that in the case of LOFAR the leakage terms are at worst a few percent, and this is further reduced due to the fact that there are many thousands of elements, which are physically identical, so any leakages would average out over the array. Also, as the orientation of the LOFAR dipoles is at 45 degrees to the north-south orientation and our sources were all observed close to transit,  both sets of dipoles would have received approximately the same amount of radiation from the source. Lastly, we note that the polarised fraction is relatively low, and that the Faraday rotation in the ISM is large at these frequencies. In general, with the bandwidths used to make pulse profiles (typically $\sim12$~MHz, see Section~\ref{sec:analysis}), the plane of polarisation is rotated through at least 180\degr in each profile, thereby further reducing the effect of polarisation calibration terms. We are therefore confident that the results presented here are not affected significantly by polarisation calibration.


\section{Analysis}
\label{sec:analysis}

The data were converted into PSRFITS format and processed using the PSRCHIVE software suite \citep{hvm04}. Regular timing observations from the Lovell telescope at Jodrell Bank Observatory \citep{hlk+04} were used to derive accurate values for the spin period ($P$) and spindown ($\dot{P}$) for the day of our wide-band observing campaign. Initial DM estimates were taken from \citet{hlk+04} (see Table~\ref{tab:psrobs} for the values used). The data were dedispersed, folded and terrestrial interference signals were removed by hand using the RFI-removal software, {\sc pazi}\footnote{A tool from the PSRCHIVE suite \citep{hvm04}. http://psrchive.sourceforge.net/ \label{fn:psrchive}}. 

Templates were made for the observations at each telescope by completely collapsing the data in both time and frequency to make an average pulse profile. The profiles were then fitted with von Mises functions\footnote{A von Mises function is given by $f(x)=\frac{e^{\kappa\cos(x-\mu)}}{2\pi I_0(\kappa)}$, where $\mu$ and $1/\kappa$ are analogous to the mean and the variance in a normal distribution. $I_0$ is the modified Bessel function. They are used by {\sc paas} because they are needed to deal with pulsars which have broad pulse profiles, although for the pulsars in our sample, which have narrow pulse profiles, (the pulse width in all cases is $<20$\% of the pulse period) they are almost identical to Gaussians.} to create analytic templates using {\sc paas}\footnotemark[3]. 

A template was created for each observing band and these templates were subsequently aligned by eye with {\sc pas}\footnotemark[3] using either the brightest peak (for PSR B0329+54) or the midpoint between the two brightest components (for pulsars B0809+74, B1133+16 and B1919+21) as the fiducial point as described in \cite{cra70}. The templates were cross-correlated with the data to get times of arrival (TOAs) using {\sc pat}\footnotemark[3] and barycentred using {\sc tempo2}\footnote{ {\sc tempo2} is a pulsar timing package for barycentring and modelling TOAs \citep{hem06}. http://www.atnf.csiro.au/research/pulsar/tempo2}. These TOAs were subtracted from a model of the pulsar's rotation in order to produce `timing residuals', which are plotted in the left hand panel of Figure \ref{fig:resid}. 

Currently, it is not possible to perform absolute timing with LOFAR. The clock corrections between the stations are known to better than a few nanoseconds, but the offset between the LOFAR clocks and Coordinated Universal Time (UTC) is not logged. This introduces time delays which are on the order of a few milliseconds between LOFAR data and the data from the other telescopes in the observations. The delays presently require an arbitrary phase offset (`jump') to be removed between data sets from different telescopes. In the case of the 8.35~GHz Effelsberg observation, where no frequency resolution was stored (see Table~\ref{tab:observations}), the determination of this phase offset uniquely defines the residual with respect to the other observations, implying that no further timing information can be derived from the observation. We have therefore omitted the Effelsberg observations from the timing analysis (though they were still used in the analysis of profile evolution, see Section~\ref{sec:prof}).

\begin{figure}[t]
\centering
\includegraphics[width=\linewidth]{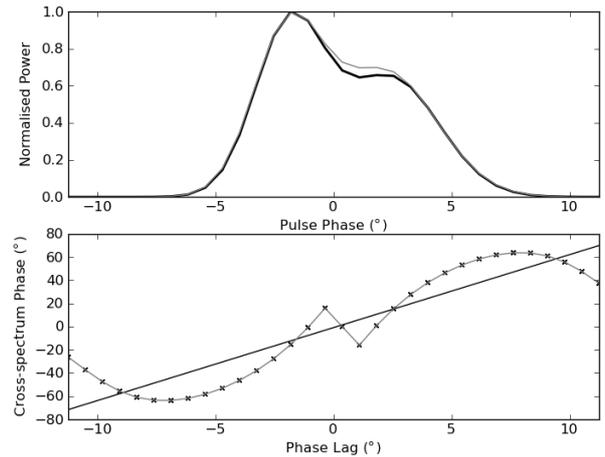}
\caption{Errors in timing PSR B1919+21. The top panel shows the HBA template (black line) and the 145~MHz data (grey line). The bottom panel shows the cross-spectrum phase of the template and the profile and a straight line fit to the data (solid line). The subtle difference between the shape of the pulse profile and the template causes a gradient in the cross-correlation phase shifting the apparent TOA.}
\label{fig:profile_evo_error}
\end{figure}

One can see in Figure \ref{fig:resid} (left) that for each of the pulsars there is clear structure in the timing residuals as a function of frequency. Often, this structure shows different slopes for each of the bands, indicating that the deviations from a good fit are not simply caused by an incorrect dispersion measure. Initially it seemed likely this might be caused by deviations from the simple form of the cold plasma dispersion relation. 

Consequently, we tried fitting the data using power laws with exponents between $-5$ and $+5$, but none of the fits significantly improved upon the chi-squared obtained from only fitting for the dispersive delay (for example, in the case of PSR B1919+21, the best power-law fit only decreased the reduced chi-squared from 1.47 to 1.44, which, given the number of degrees of freedom in the model, still corresponds to a $\sim99\%$ chance that there is unmodelled structure in the residuals). In some cases, the residuals also showed structure which could not be explained by a single function (for example, a positive gradient in the LBA data, a negative gradient in the HBA data and a positive gradient in the data from Jodrell Bank). No simple power law can account for the different slopes in our timing residuals, so we conclude that, the structure in the residuals is not caused by the ISM or aberration and retardation in the pulsar magnetosphere.

The real cause of the systematic errors is the pulse profile changing as a function of frequency, as noted in \cite{amg07}. As the profile changes, the template which is used in the cross-correlation becomes more and more inaccurate and there is a systematic drift in the residuals towards earlier or later TOAs. Figure~\ref{fig:profile_evo_error} shows how these errors arise. The top panel shows the template used for the HBA observation of PSR B1919+21 along with the observed pulse profile at 145~MHz. The bottom panel shows the cross-spectrum phase \citep[see][for details]{amg07} of the two profiles and a straight line fitted to the data. There is a gradient in the fitted line which means that the peak of the amplitude of the cross-correlation (used to produce the TOAs) is shifted slightly, causing the apparent TOA to be delayed or advanced compared to its true value.

\begin{figure*}
\begin{minipage}{0.5\linewidth}
\centering
\includegraphics[height=0.2\textheight]{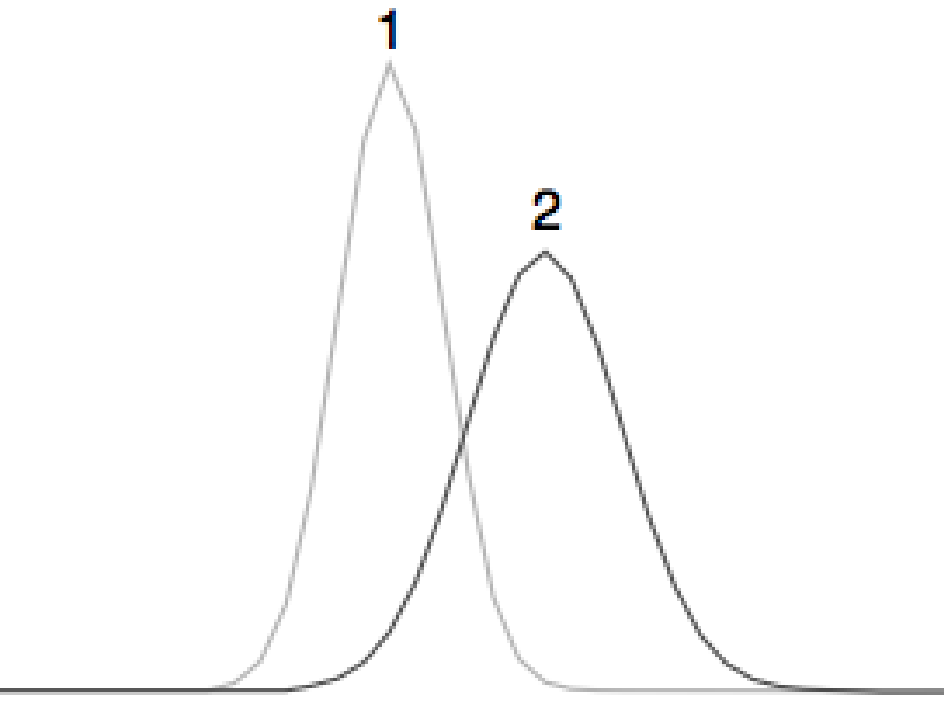}
\includegraphics[height=0.5\textheight, trim=2cm 2cm 7cm 8cm, clip=true]{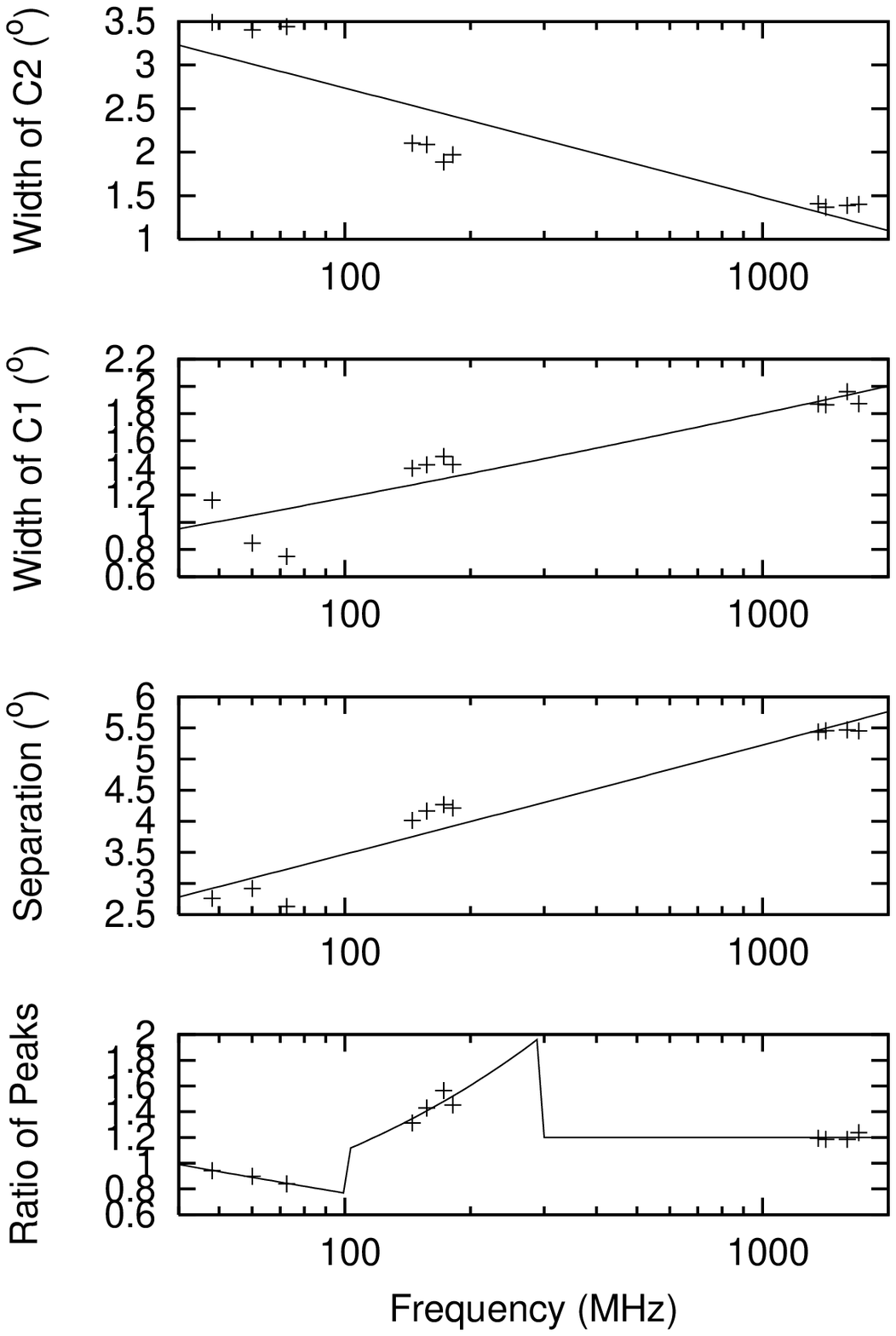} 
\end{minipage}
\hspace{0.2cm}
\begin{minipage}{0.5\linewidth}
\centering
\includegraphics[height=0.75\textheight, trim = 1cm 1cm 1cm 2.5cm, clip = true]{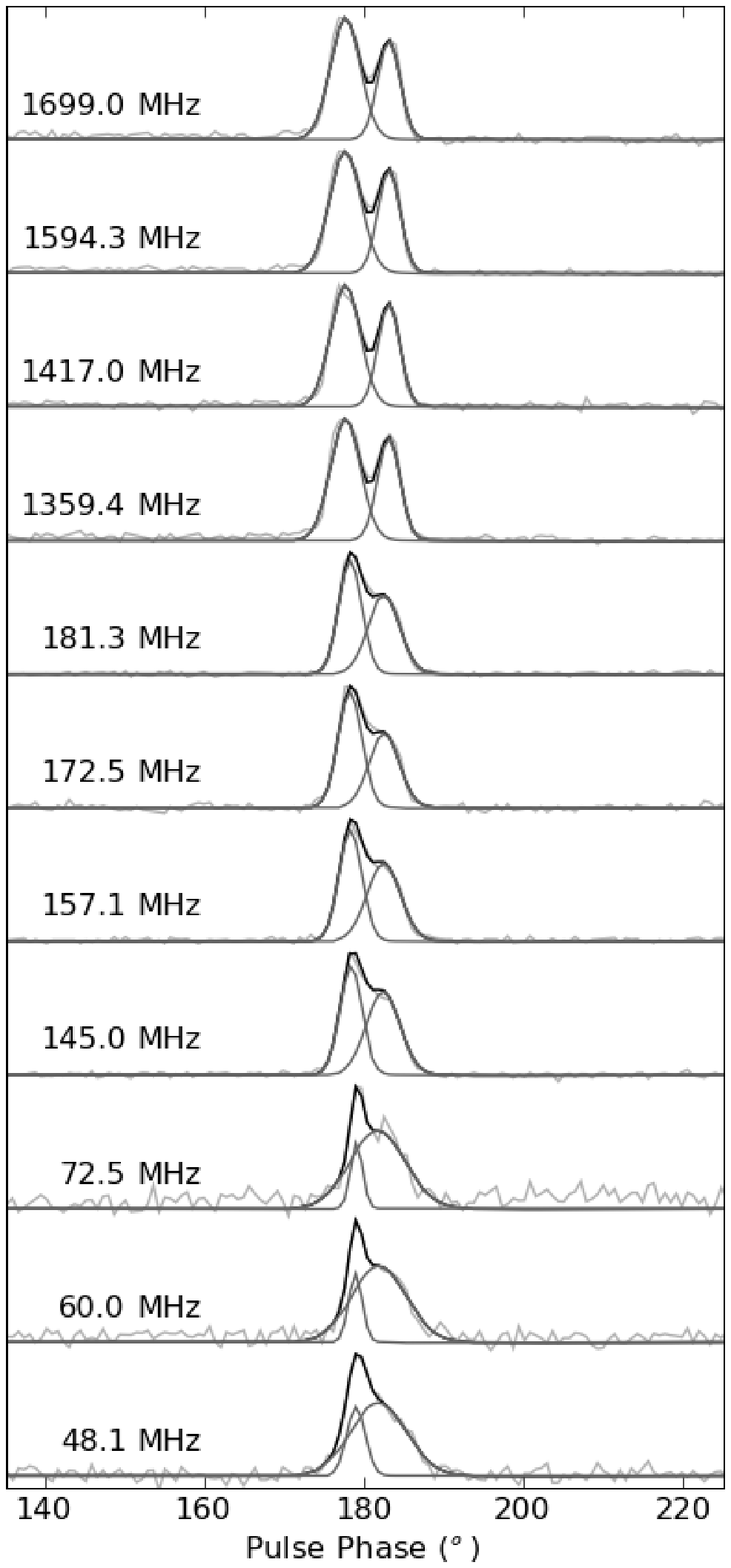} 
\end{minipage}
\begin{minipage}{0.1\linewidth}
\end{minipage}
\caption{PSR B1919+21 is the simplest of our models, a double peaked pulsar. The top left panel shows the model used for the pulsar, two Gaussians with the fiducial point set as the midpoint between the peaks of each component. The Gaussians are fitted to the data using a least squares fitting algorithm, allowing their width, height and separation to vary. This fitting is shown in the right hand panel. The data is plotted in light grey, the two fitted components are plotted in dark grey, and the sum of both fitted components is plotted in black. The Gaussian parameters for each of the observations are recorded and plotted as a function of frequency in the bottom left panel. These parameters are then fitted with power laws to get a model of the pulse profile as a function of frequency. This global model is subsequently used to produce templates for cross-correlation. The `ratio of peaks' plot has discontinuities because different power laws were fitted in each observing band. PSR B1919+21 was not detected in the Effelsberg observations as the source was too weak.}
\label{fig:Fitting}
\end{figure*}

\begin{figure*}
\includegraphics[width=\linewidth, trim=0cm 9.5cm -4cm 7cm, clip=true]{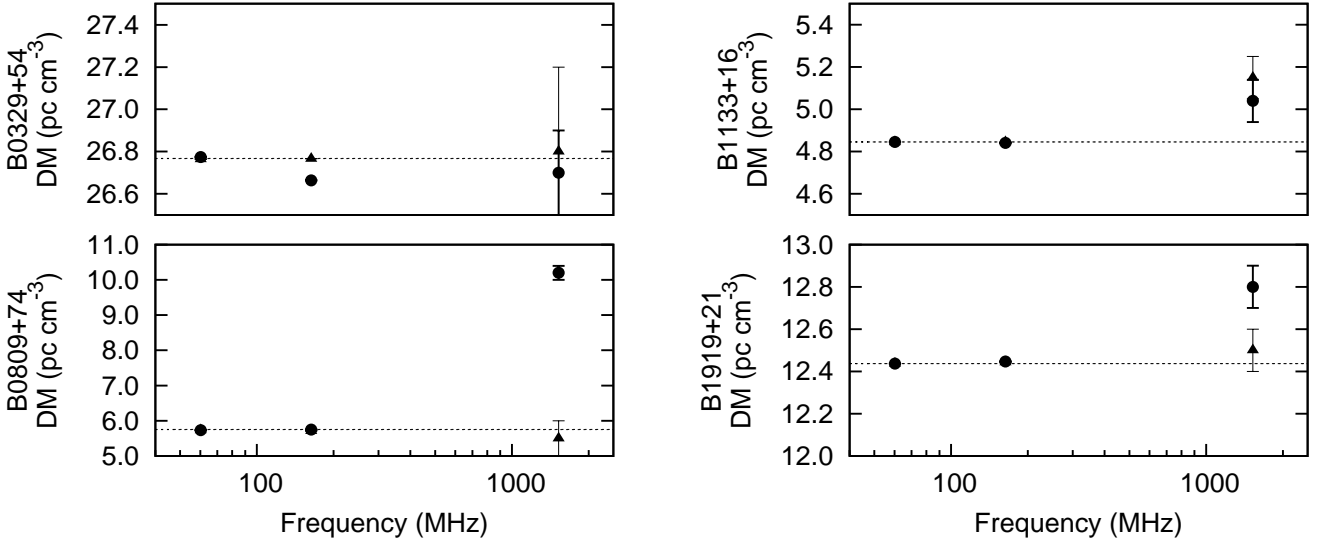}
\caption{A plot of frequency against apparent DM in each of the observing bands. The circluar points show the DM obtained using a static template, and the triangular points show the DM from a frequency-dependent template based on our models of the pulsars' pulse shape evolution. In each case (apart from that of PSR B1133+16) the model-based template provides a consistent DM across all frequency bands, while use of the static template often results in significantly different DM values.}
\label{fig:dm_var}
\end{figure*}

\begin{table*}
\caption{Apparent DM in each frequency band of our observations using a static template and a template based on a frequency-dependent model. DM$_\mathrm{avg}$ is the DM using all of the observing bands.}
\centering
\label{tab:dm_var}
\begin{tabular}{cccccc}
\hline
		&	& DM$_\mathrm{LBA}$ 	& DM$_\mathrm{HBA}$ 	& DM$_\mathrm{Jod}$  & DM$_\mathrm{avg}$	\\
\hline
B0329+54 & static & 	$26.774\pm 0.001$	&  $26.664\pm0.0003$	 & $26.7\pm0.2$ &	$26.7673\pm0.0002$ \\
		 & model & 	$26.764\pm 0.001$	&  $26.7662\pm0.0004$	 & $26.8\pm0.4$ &	$26.7641\pm0.0001$ \\
\hline
B0809+74 & static & 	$5.735\pm0.006$	& 	$5.750\pm0.007$ 	&  $10.2\pm0.2$ & $5.752\pm0.001$	\\
		& model & 	$5.735\pm0.005$	& 	$5.71\pm0.01$	 	&  $5.5\pm0.5^a$ &	$5.733\pm0.001$	\\
\hline
B1133+16 &static & 	$4.8456\pm0.0003$	& 	$4.8412\pm0.0005$	& $5.04\pm0.1$ & 	 $4.8459\pm0.0001$	\\
		& model & 	$4.8450\pm0.0003$	& 	$4.8446\pm0.0005$	& $5.15\pm0.1$ & 	$4.8451\pm0.0001$ \\
\hline
B1919+21 & static & $12.4370\pm0.0001$	& $12.447\pm0.001$ 	& $12.8\pm0.1$ & 	$12.4373\pm0.0001$ \\
		 & model & $12.4371\pm0.0003$	& $12.443\pm0.002$ 	& $12.5\pm0.1$ & 	$12.4370\pm0.0001$ \\
\hline
\end{tabular}
\tablefoot{
\tablefoottext{a}{DM value computed with step in residuals removed.}
}
\end{table*}

This effect has only been brought to attention recently by \cite{amg07}, because pulsar timing is normally done at high frequencies (where the pulse shape changes are less obvious), and with modest bandwidths. With these narrow bandwidths, it is easy to incorrectly interpret this structure as an incorrect DM.  With the wide fractional bandwidth and high frequency resolution of our simultaneous data, however, the effect cannot be fitted with a simple $\nu^{-2}$ law, and so the structure is more easily identified.

To remove the effects of profile evolution and find the true DM value and residuals, we first modelled how the pulse profiles evolve as a function of frequency. The data were divided into narrower-frequency subbands. Each subband was chosen such that the profile evolution within it was not significant, but there was still enough bandwidth so that the pulsar was detected clearly. The size of these bands depended on the signal-to-noise ratio of the observation, but typically we used 4 subbands per telescope, corresponding to $\sim12$~MHz in the LOFAR data and $\sim100$~MHz in the Lovell data. Each subband was then collapsed in frequency and folded in time to create a pulse profile, which, unlike the previous analysis, we fitted with Gaussians\footnote{As noted earlier, for these pulsars Gaussian functions are almost identical to von Mises functions, so the different functions used for fitting the static and model-based templates has no effect on the timing residuals.}. 

The width, height and positions of the Gaussians were free to vary, and the best fit values for each of them were recorded and used to create a continuous model of how the pulse profile evolves as a function of frequency. The fiducial point was chosen as either the centre of the main pulse or the midpoint of two brightest components depending on the morphology of the pulse profile \citep[see][]{cra70}.

This process is shown for PSR B1919+21 in Figure \ref{fig:Fitting}. We used power laws to model the frequency-evolution of each of the fitted parameters,  as they are the simplest functions which fit the data well. However, as can be seen in Figure~\ref{fig:Fitting} the fits were not perfect. The separation of the peaks follows a power law well (although the exponent is positive, contrary to what is expected by radius-to-frequency mapping, see Section~\ref{sec:B1919+21}), but the widths of the components are not well described by power laws. This is due to the overlap between components. In the overlapping region it is unclear how much power belongs to each component, and the solution which is achieved from the fit is not unique. The ratio of the peaks also could not be fitted with a single power law, which was the case for all of the pulsars in our sample. Instead we used a power law for each frequency band. This is because both of the components have different spectral profiles (which may contain one or more breaks) and as we don't have absolute flux values, it is unclear whether one component is getting brighter or the other is getting fainter. Scintillation across the different frequency bands could also contribute slightly to the discontinuities in the component amplitudes, although the effects of scintillation will be small in the average pulse profiles because of the long integration times used in these observations.

\begin{figure*}[t]
\centering
\includegraphics[width=\linewidth, trim=1cm 2cm -6cm 17cm, clip = true]{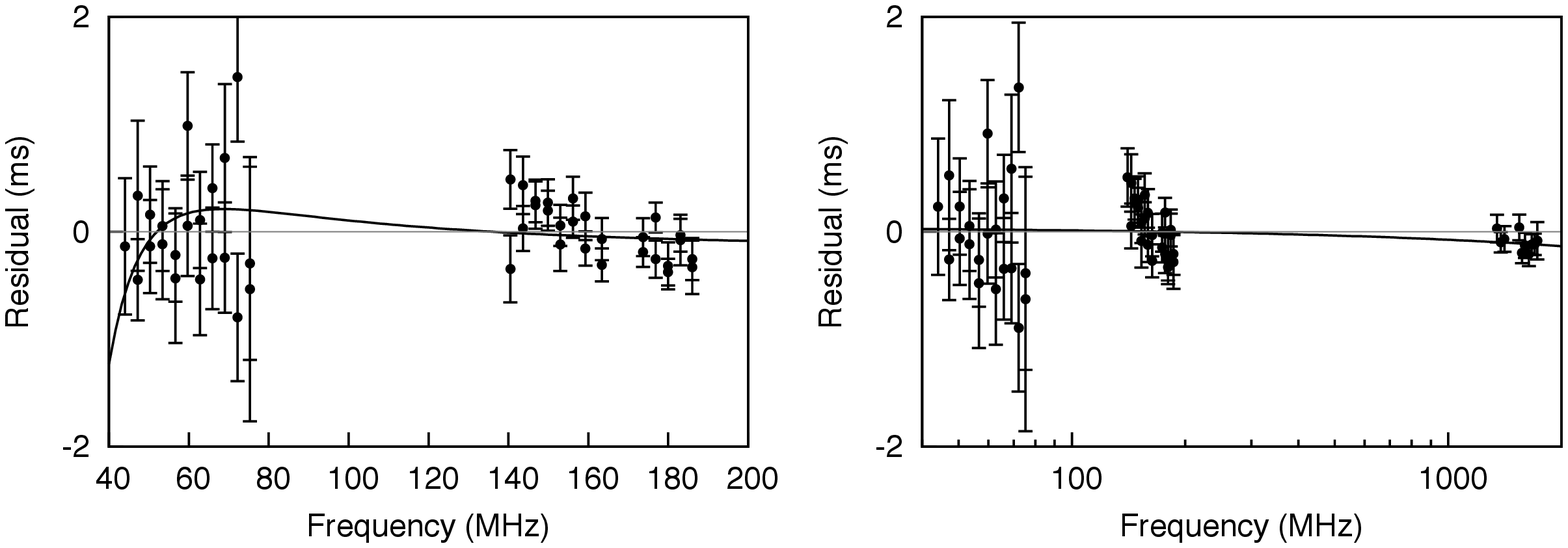}
\caption{Simulations of structure in our residuals of PSR B1919+21. The top panel shows an example of our TOAs, with a simulated $\nu^{-4}$ ISM-like delay, which would be 0.84~ms at 48~MHz, added to them. The fit to the data is shown by the black line, and the null hypothesis (no ISM delay) is plotted in grey. Similarly, the right hand panel shows an example of a simulated aberration/retardation-like delay with a $\nu^{0.6}$ dependence which is 0.28~ms at 180~MHz. Because the errors on the TOAs are much smaller at high frequencies, we are much more sensitive to delays at high frequencies, despite the fitted jump. Note that the sensitivity of both the LOFAR LBA and HBA observations is now vastly improved and so we should be able to better constrain (or even detect) some of these effects in the near future.}
\label{fig:sims}
\end{figure*}

Our model of the pulsar was then used to create a template for each frequency channel, which was cross-correlated with the data to get TOAs using {\sc pat} in the same way as the single templates were. 

The right hand side of Figure \ref{fig:resid} shows the residuals obtained from using the templates derived from this frequency-dependent model. Using these model-based templates for timing improves the residuals and reduces the systematic errors which were seen in the initial residuals. With the frequency-dependent templates the residuals show far less systematic trends and agree to within the error bars on a single DM in most of our observations (see Figure~\ref{fig:dm_var} and Table~\ref{tab:dm_var}). PSR B1133+16 seems to show some structure in its residuals. Its DM at $\sim1400$~MHz is more than 3 sigma from its DM at low frequencies and this cannot be removed by using a frequency-dependent template. This is possibly due to the finite sampling of the data and template which means that at the highest frequencies the pulse profile of PSR B1133+16 (whose components are the narrowest of all the pulsars in our sample) is barely resolved. Subsequent observations taken at Jodrell Bank, with higher time resolution, show no such structure. None of the other pulsars in our sample were affected by this problem.

We will return to the issue of the component evolution in pulsars B0329+54, B0809+74, B1133+16 and B1919+21 and its impact on pulsar timing (Section \ref{sec:prof}), but first we will consider the influence of the ISM.

\section{Simulations of Non-dispersive Delays}
\label{sec:sims}

Before we started looking for extra structure, the data had already been fitted for DM and a phase offset between LOFAR and the Lovell telescope. The size of the error bars from the cross-correlation are also very different in each of our observing bands. For these reasons, it was necessary to perform simulations to determine how sensitive the data truly are to these frequency-dependent effects.

To simulate the effects of the ISM, we added a delay to the folded profile in each subband of our data according to a  $\nu^{-4}$ scaling law (most second-order ISM delays scale roughly as  $\nu^{-4}$, see Section~\ref{sec:ism}). This data was then cross-correlated with a frequency-dependent template, fitted for dispersion measure and a jump between the LOFAR data and the L-band data, and displayed in the same way as the real data to produce residuals which had an artificial delay added to them. 

We used the least squares method to fit the residuals with a $\nu^{-4}$ power law. The chi-squared value of the fit and the number of degrees of freedom were computed, and used to compare the likelihood that the structure in the residuals was caused by a $\nu^{-4}$ delay with the likelihood that it was caused by chance\footnote{This was done numerically, using the chi-squared calculator, see http://www.fourmilab.ch/rpkp/experiments/analysis/chiCalc.html}. We reduced the magnitude of the delay which was introduced to the data until the residuals had an equal probability of being caused by chance as being caused by an ISM-like power law. This delay was set as the upper limit on the magnitude of this effect in our timing residuals. This process is demonstrated in the left panel of Figure~\ref{fig:sims}. From these simulations, we determined that our sensitivity to steep, negative, frequency-dependent power laws is dominated by the error bars associated with the LBA observations.

In the same way, we also simulated the effects of an aberration/retardation-like delay. \cite{cor78} showed that the emission height, $r(\nu) \propto \nu^{-2x}$, where $x$ is a constant which depends on the radius-to-frequency mapping of the particular pulsar. The author then used data to show that typically $x$ falls in the range $0.21-0.55$. 

The delay in the pulse arrival time is proportional to the height at which it is emitted (see Equation~\ref{eq:abret}). From this, we can deduce that $\Delta t$ should scale with $\nu^{0.4-1}$. For our simulations, we elected to use the average value of $x$ found by Cordes, so chose a power law which scales with $\nu^{0.6}$.

Again, we compared the likelihood of a fitted function and no function, and reduced the magnitude of the effect until it was undetectable in our timing residuals (see the right panel of Figure~\ref{fig:sims}). Remarkably, the fitted jump has very little effect on our ability to detect aberration and retardation effects, and because the error bars in our high frequency data are much smaller than those at low frequencies we are, in fact, much more sensitive to high frequency delays. If the power law is steeper than $\nu^{0.6}$ our sensitivity to these effects increases. Table~\ref{tab:sims} shows the determined upper limits on high and low frequency delays in our data. 

\begin{table}
\centering
\caption{Upper limits derived from simulations.}
\begin{tabular}{ccc}
\hline
PSR 			& $\nu^{-4}$ delay (ms)	& $\nu^{0.6}$ delay (ms) 	\\
			&	at 48~MHz		&   at 180~MHz			\\ \hline
B0329+54	&	1.95				&		0.65			\\
B0809+74	&	3.84				&		1.28			\\
B1133+16	&	1.05				&		0.35			\\
B1919+21	&	0.84				&		0.28			\\
\hline
\end{tabular}
\label{tab:sims}
\end{table}%

\section{ISM Effects}
\label{sec:ism}

\subsection{Impact on Pulsar Timing}

\begin{figure}[b]
\centering
\includegraphics[width=\linewidth, trim=0.5cm 0cm 1.5cm 1cm, clip=True]{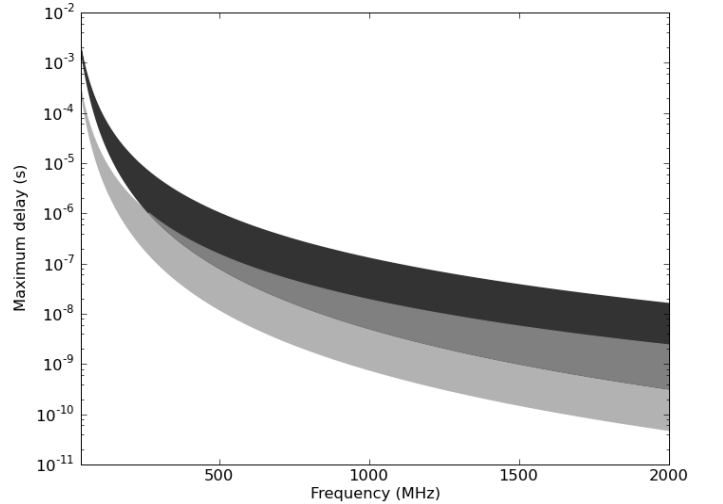}
\caption{These curves show the upper limits on the size of delays from second-order ISM effects extrapolated up to higher frequencies. The dark grey area is for PSR B0809+74, which had the largest RMS residuals in our timing fits; the light grey area is for PSR B1919+21, which had the smallest RMS residuals. They are scaled with $\nu^{-3}$ and $\nu^{-4}$ which are the lower and upper bounds for scaling of the second-order ISM effects.}
\label{fig:extrapolate}
\end{figure}

There are a number of second-order effects caused by the ISM which could induce additional delays in our data. These effects scale strongly with wavelength and are at their strongest at low frequencies. As shown by our simulations, the lack of any remaining structure in the residuals indicative of unmodelled effects shows there is no significant indication of super-dispersion, refraction, anomalous dispersion or frequency-dependent DM variations. We can, however, still use these observations to place important limits on the magnitude of some of these effects, at least for these four sources.

Taking the maximum unmodelled (i.e. non-dispersive) ISM delay in the LBAs and extrapolating it to higher frequencies gives an indication of the impact of the ISM on pulsar timing projects like Pulsar Timing Arrays \citep[PTAs, see][]{rom89,fb90}. Figure \ref{fig:extrapolate} shows the magnitude of ISM effects as a function of frequency for two pulsars with the largest (B0809+74) and smallest (B1919+21) deviations from white noise in our sample. The largest possible delay in our data (see Table~\ref{tab:sims}), scaled with the gentlest gradient (i.e. the largest possible deviation) still corresponds to only $\sim50$~ns at 1400~MHz. Although this figure will change significantly for every line-of-sight, depending on the parameters of the ISM (and potentially when the observation was taken) this upper limit is only roughly half of the $\sim100$~ns precision required for the first generation of PTAs \citep{jhlm05}. 

\citet[][and references therein]{cs10} describe a number of frequency-dependent delays, which are caused by the distribution and number of free electrons along the line-of-sight. Combining these relations with our timing residuals allows us to constrain some properties of the ISM (see Table~\ref{tab:limits}).

\subsection{Frequency-dependent DM Variations}
\begin{figure}
\centering
\includegraphics[width=\linewidth, trim=0cm 0cm 0cm 15cm, clip=True]{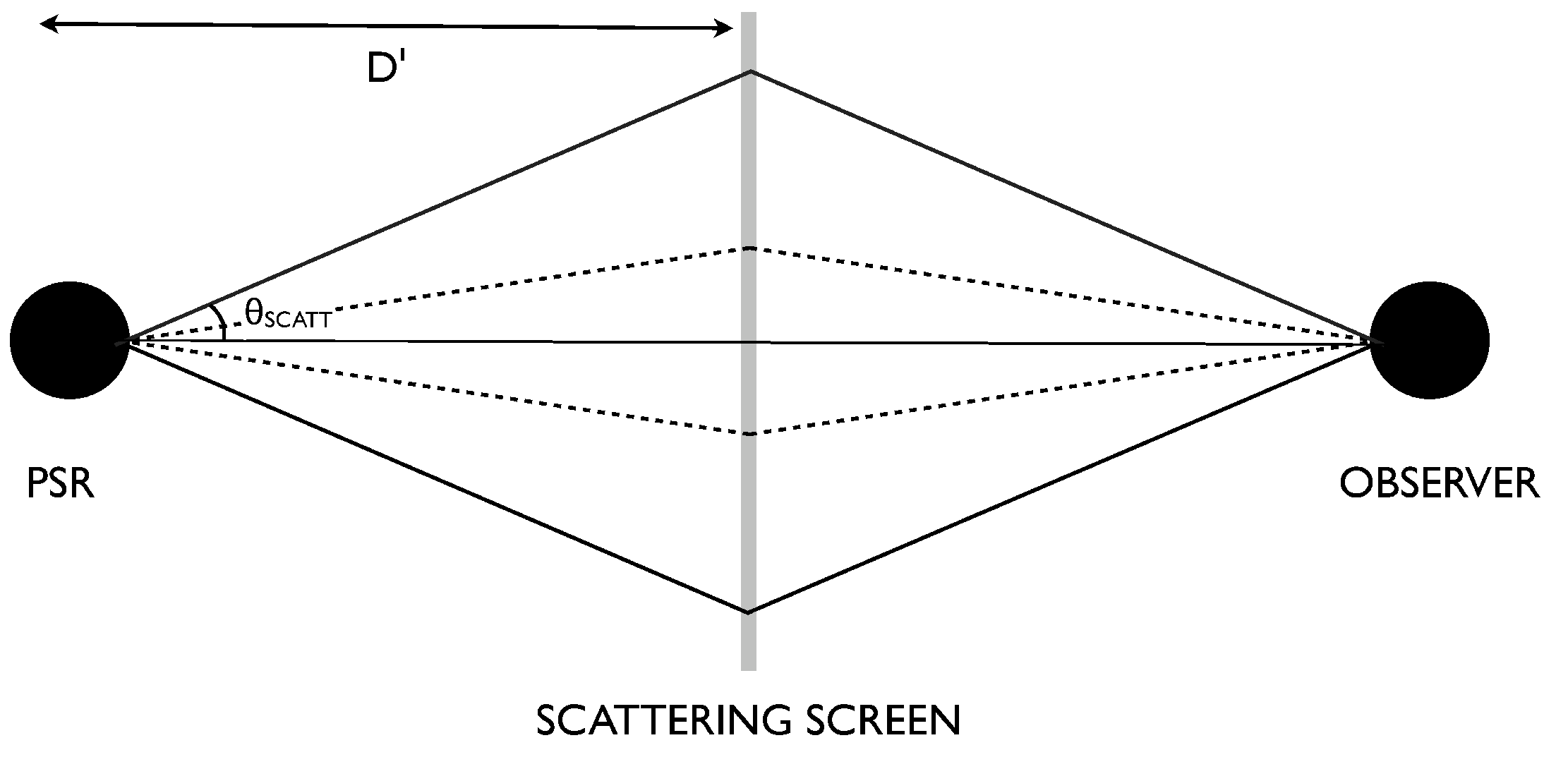}
\caption{The size of the scattering cone is larger at low frequencies (dark line) than at high frequencies (dashed line). The DM is a flux-weighted average over the area of the scattering disk so it varies with frequency as the scattering cone changes size.}
\label{fig:multipath}
\end{figure}

Frequency-dependent DM variations are caused by scattering by the ISM. The diameter of the scattering disk (the region in the sky where scattered radiation is received from) increases as $\sim\nu^{-2.2}$. The DM, which is an average over the scattering disk weighted by the flux received from a given direction, therefore also varies as a function of frequency (Figure~\ref{fig:multipath}). 

If we assume that all of the scattering material is concentrated in a thin screen half-way along the line-of-sight, the variation in DM is given by \citep{cs10}:
\begin{equation}
\delta \mathrm{DM} = 2.82\times10^{4}D'^{5/6}\nu^{-11/6}\mathrm{SM}~\mathrm{pc~cm^{-3}},
\end{equation}
where $D'$ is the distance between the pulsar and the scattering material (in kpc), $\nu$ is the observing frequency in MHz and SM is the Scattering Measure along the line-of-sight in units of kpc~m$^{-20/3}$. This corresponds to an RMS variation in time delay of:
\begin{equation}
\Delta t_{\mathrm{DM},\nu} = 3.79\times10^{4} D'^{5/6}\nu^{-23/6} \mathrm{SM}~\mathrm{s}.
\end{equation}
This allows us to constrain the SM along the line-of-sight (see Column 4 of Table \ref{tab:limits}). The SM must be fairly large for this effect to be detectable and all of the pulsars which we observed are at low DM ($<30$~pc~cm$^{-3}$) and show very little scattering (with the exception of PSR B0329+54, no scattering is visible in any of the pulse profiles). Our upper limits are about 2--3 orders of magnitude above those from the NE2001 model of \cite{cl02}. 

To detect this effect in PSR B0329+54 (assuming the SM from NE2001) would require a timing residual RMS of 10~$\mu$s, so it is not surprising that this effect was not detectable in our data. However, it is possible that this effect could be detected in the future with LOFAR using pulsars with a higher SM (for example PSR J2044+4614 or PSR B2036+53, both of which are closer to the Galactic plane). The LOFAR data presented here is far from full sensitivity, and, when the data are fully calibrated, and all 24 of the stations in the LOFAR core can be combined coherently, the telescope's sensitivity will be improved by at least an order of magnitude. 


\subsection{Refraction}
\begin{figure}
\centering
\includegraphics[width=\linewidth, trim=0cm 0cm 0cm 1cm,clip=True]{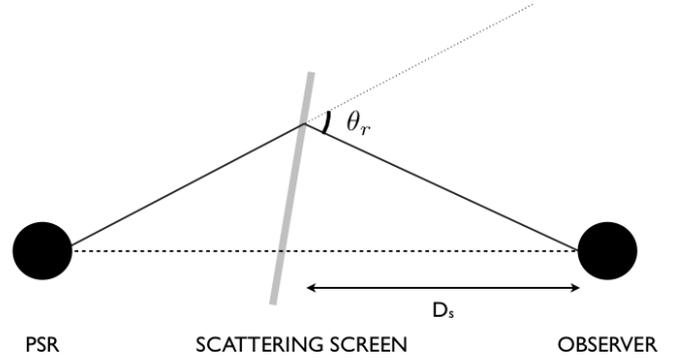}
\caption{A gradient in the electron density perpendicular to the line-of-sight causes rays to be bent. The size of the angle which the rays are bent through depends on the frequency of the radiation. Low frequencies (dark line) are bent through a larger angle and so must travel along a longer path to reach the observer, delaying them with respect to higher frequencies (dashed line).}
\label{fig:refraction}
\end{figure}

Refraction can also introduce a delay into pulsar timing data at low frequencies. If there is a gradient in the interstellar electron density perpendicular to the line-of-sight, rays of light are refracted, and bent through some angle $\theta_r$. For a thin screen this is given by \cite{cfl98}:
\begin{equation}
\label{eq:ref1}
\theta_r = \frac{\lambda^{2}r_e}{2\pi}\frac{d}{dx} N_e(x),
\end{equation}
where $r_e$ is the classical electron radius and $N_e$ is the column density of electrons along the line-of-sight. At lower frequencies, light is bent through a larger angle. The larger the angle, the further the light must travel and the longer it takes to arrive at the observer (see Figure \ref{fig:refraction}). 

\cite{cs10} showed the delay between a straight path and a refracted path is given by:
\begin{equation}
\label{eq:ref2}
\Delta t_\mathrm{ref} = \frac{1}{2c}D_\mathrm{eff}\theta_r^2 .
\end{equation}
where $D_\mathrm{eff}$ is a characteristic distance, given by:
\begin{equation}
D_\mathrm{eff} = \frac{D-D_s}{D_s/D}~,
\end{equation}
$D$ is the distance between the observer and the pulsar and $D_s$ is the distance between the observer and the scattering screen. Assuming the line-of-sight is dominated by one screen (which is a reasonable assumption given the evidence in \citealt{sti06} and \citealt{bmg+10}, and that our sample is quite nearby), the thin screen model is valid. By substituting $\theta_r$ from Equation~\ref{eq:ref1} into Equation~\ref{eq:ref2} and rearranging for $\nu$ we can derive an expression for the time delay between refracted and unrefracted rays:
\begin{equation}
\Delta t_{\mathrm{ref}} = 3.47\times10^{14}\frac{D_\mathrm{eff}}{\nu^4}\left(\frac{d}{dx} \mathrm{DM}\right)^2~\mathrm{s},
\end{equation}
where $\frac{d}{dx}\mathrm{DM}$ is the average gradient in the DM perpendicular to the line-of-sight  in pc~cm$^{-3}$~AU$^{-1}$, $\nu$ is in MHz and $D_\mathrm{eff}$ is in kpc . This number can be constrained from the lack of structure in our timing residuals and can be used as an indicator of how much the DM of a pulsar is likely to change. 

It follows from the derivation above, that the $\frac{d}{dx}\mathrm{DM}$ value obtained holds over distances $d x$ of the size of the scattering cone. At these frequencies the scattering cone for a typical (nearby) pulsar is on the order of a few AU (for B0329+54 the scattering cone is 10~AU at half the distance to the pulsar), so the most natural units to use for this quantity are pc~cm$^{-3}$~AU$^{-1}$. The unit pc~cm$^{-3}$~AU$^{-1}$ also corresponds to how much the DM of a pulsar at a distance $D$ travelling 250~km~s$^{-1}$ will change in approximately one week. It should be noted however, that the number is an average over the entire scattering disk, and so is not sensitive to small scale anisotropies. Upper limits are shown for $\frac{d}{dx} \mathrm{DM}$  for the pulsars in our sample in Column 5 of Table \ref{tab:limits}.

The main source of error in this number arises from $D_\mathrm{eff}$ as the distance to the scattering screen is unknown. The function is very sensitive to nearby scattering screens ($D_s< 0.25D$) but is not very sensitive to distant scattering screens ($D_s>0.75D$).  It is, however, approximately constant for $0.25 < D_s < 0.75$, varying by only a factor of $\sim3$ in this range.  In our calculations, we assume that the scattering screen is roughly half way along the line-of-sight and $D_\mathrm{eff} \approx D$.

\subsection{Anomalous Dispersion}
By modifying Equation~\ref{eq:disp} to include gyro-motion in a magnetic field and electron collisions, and including the quartic term in the Taylor expansion of $1/v_g$, we can write a more general version of the plasma dispersion law, which can be expressed as the sum of three functions \citep[Equations \ref{eq:t1}, \ref{eq:t2} and \ref{eq:t3}, see][]{pw92}:
\begin{equation}
\Delta t_{\mathrm{DM}} = \Delta t_{\gamma_1} + \Delta t_{\gamma_2} + \Delta t_{\gamma_3}.
\end{equation}
The $\gamma_1$ term: 
\begin{equation}
\Delta t_{\gamma_1} = \frac{\mathrm{DM}}{2.41\times10^{-4}\nu^{2}} (1+\beta_\mathrm{therm}^2)~\mathrm{s},
\label{eq:t1}
\end{equation}
is the normal dispersive delay term multiplied by an additional factor which depends on the ratio of the thermal velocity of the electrons to the speed of light, $\beta_\mathrm{therm}$:
\begin{equation}
\beta_\mathrm{therm} = \sqrt{\frac{kT}{mc^2}}~.
\end{equation}
Where $k$ is the Boltzmann constant, $T$ is the temperature of the plasma and $m$ is the mass of an electron. This means that if the temperature is non-zero it adds a small contribution to the dispersive delay. In practice this is a very small number and is absorbed into the $\nu^{-2}$ law making the DM slightly higher than it would be normally, but not altering the power law of the delay. Even an unrealistic change in temperature as large as 100~K only modifies the apparent DM by 0.01\%, and would be indistinguishable from the normal dispersive delay.

\begin{table*}
\caption{Derived values of the dispersion measure (DM), and upper limits on the scattering measure (SM), perpendicular gradient in dispersion measure ($\frac{d}{dx} \mathrm{DM}$) and emission measure (EM) derived from the analysis described in the text. The values of DM are shown here only for comparative purposes, and correspond to the average DM derived earlier from the frequency-dependent model. For more information, see Table~\ref{tab:dm_var}.}
\centering
\begin{tabular}{cccccc}
\hline	
 	    	 & Delay & DM                     & SM                   & $\frac{d}{dx} \mathrm{DM}$ & EM   		\\ 
		 & (ms)    &  (pc cm$^{-3}$)&(kpc~m$^{-20/3}$) & (pc~cm$^{-3}$~AU$^{-1}$)         &(pc cm$^{-6}$)\\ 
\hline
B0329+54 & 1.95    &    26.764	 	 & $<0.25$             &    $<5.3\times10^{-6}$ & $<42000$ \\
B0809+74 & 3.84    &      5.733            & $<1.02$             &    $<1.2\times10^{-5}$ & $<82000$ \\
B1133+16 & 1.05    &      4.845 	          &  $<0.32$	     &    $<6.7\times10^{-6}$ & $<22000$   \\
B1919+21 &  0.84   &   12.437	          &  $<0.155$	     &    $<4.4\times10^{-6}$ & $<18000$   \\
\hline
\label{tab:limits}
\end{tabular}
\end{table*}

The $\gamma_2$ term is because of the plasma being weakly magnetized. For a circularly polarised wave:
\begin{equation}
\Delta t_{\gamma_2}= \frac{0.0286\mathrm{RM}}{\nu^{3}}~\mathrm{s},
\label{eq:t2}
\end{equation}
where RM $= \int_{0}^Ln_eB_{||}\mathrm{d}l$ is the Rotation Measure in rad~m$^{-2}$. This only affects circularly polarised sources and is generally negligible. The RM along a given line-of-sight is closely related to the SM and the DM, so when a pulsar's RM is large enough to make this effect detectable, scatter broadening makes the pulsar undetectable. An RM of 280 would be needed to produce a detectable delay of 1~ms at 20~MHz. The scattering time for a pulsar with this RM would be $\sim4$~s, which, in most cases, would render the pulsar undetectable (assuming $\mathrm{RM}\sim5\times\mathrm{DM}$, the average from the ATNF pulsar catalogue \citealt{mhth05}, and the empirical scattering law from \citealt{bcc+04}).


The $\gamma_3$ term depends on the Emission Measure, EM $=\int_{0}^{D} n_e^2 \mathrm{d}l$ along the line-of-sight:
\begin{equation}
\label{eq:EM}
\Delta t_{\gamma_3} = \frac{\mathrm{EM}}{4.0\nu^4}~\mathrm{s},
\label{eq:t3}
\end{equation}
where EM is given in pc~cm$^{-6}$. For a uniform distribution of plasma between the pulsar and observer the contribution from the $\nu^{-4}$ term is small even at low frequencies and is probably not detectable. However if the electrons are arranged in a clumpier distribution along the line-of-sight (as discussed in \citealt{klll08}) this term becomes larger. 

Whilst both the $\beta_{therm}$ and $\gamma_2$ terms are too small to detect, the fact that no delay is observed in our data can be used to constrain the EM along the line-of-sight through the $\gamma_3$ term. The upper limits are given in Column 6 of Table \ref{tab:limits}. Although they are currently around five orders of magnitude higher than the values predicted by NE2001 \citep{cl02},  using this method on observations at lower frequencies and with higher sensitivity, could provide a new way to measure the EM along a given line-of-sight.

\section{Magnetospheric Effects}
\label{sec:mag}
In addition to the delays caused by the ISM, there are potential sources of delay from within the pulsar magnetosphere itself. We can use the absence of any additional delays in our timing fits to constrain the composition of the magnetosphere and the emission height above the magnetic poles.

\subsection{Aberration and Retardation}
According to radius-to-frequency mapping \citep{rs75,mt77,cor78}, high frequency emission is thought to originate close to the neutron star whilst low frequency emission comes from higher in the pulsar magnetosphere. The range of emission heights at different frequencies can be obtained, independently of the radius-to-frequency mapping model, through the effects of differential aberration and retardation \citep{cor78}.

Retardation is the time delay caused by the difference in path length from the different emission sites. For emission which originates at an altitude $r_\mathrm{min} <  r_\mathrm{max}$, the time delay between the two paths is given by:
\begin{equation}
\Delta t_\mathrm{ret} = \frac{r_\mathrm{max} - r_\mathrm{min}}{c}.
\end{equation}
Aberration is caused by the co-rotation of the magnetosphere, which bends the radiation beam towards the direction of rotation and therefore causes pulses to arrive earlier than they would if they were to travel along a straight path. Aberration increases at larger radii, so the result is to delay pulses from low altitudes by:
\begin{equation}
\Delta t_\mathrm{ab} = \sin\alpha\Delta t_\mathrm{ret},
\end{equation}
where $\alpha$ is the angle between the pulsar's magnetic and rotational axes. Combining the two effects gives the total time delay for a given radius:
\begin{equation}
\label{eq:abret}
\Delta t_{A/R} = (1 + \sin\alpha)\frac{r_\mathrm{max} - r_\mathrm{min}}{c}.
\end{equation} 

In our data, assuming the fiducial points in our frequency-dependent models are aligned correctly, there is no remaining structure in the residuals that has not been successfully modelled by a quadratic frequency dependence. This can be used to constrain $\Delta R = r_\mathrm{max} - r_\mathrm{min}$, the distance between the heights at which different frequencies are emitted (see Table ~\ref{tab:abret}). 

For pulsars B0329+54 and B1133+16, which exhibit typical `conal' behaviour \citep[see for example][]{ran83}, it is also possible to use radius-to-frequency mapping to determine upper limits on the absolute height of emission from the pulsar surface. At low frequencies radius-to-frequency mapping suggests that pulse profiles are broadened as the star's dipolar magnetic field lines move further apart high in the magnetosphere. For a neutron star with a dipolar magnetic field the ratio of the widths of the profiles ($\theta_\mathrm{\nu2} > \theta_\mathrm{\nu1}$) at two frequencies ($\nu_1 > \nu_2$) can be related to the emission heights by the equation \citep{cor78}:
\begin{equation}
\frac{\Delta R}{r_\mathrm{min}} = \left(\frac{\theta_\mathrm{\nu2}}{\theta_\mathrm{\nu1}}\right)^2 -1.
\end{equation} 
This equation can be used in conjunction with the values of $\Delta R$ derived earlier to determine  upper limits on $r_\mathrm{max}$, the absolute height of the 40~MHz emission. This analysis was not performed for pulsars B0809+74 and B1919+21, as our observations of their pulse profile evolution do not agree with the standard picture of radius-to-frequency mapping (see Section~\ref{sec:prof}). 

\begin{table}
\caption{Constraints on emission heights for frequencies between $\sim40$ and $\sim180$~MHz from aberration/retardation arguments. The $\alpha$ values are taken from \cite{lm88}. $r_\mathrm{max}$ was not calculated for pulsars B0809+74 and B1919+21 as the frequency evolution of their pulse profiles is not consistent with a dipolar magnetic field and simple radius-to-frequency mapping.}
\centering
\begin{tabular}{ccccc}
	\hline	
	 	    	 & $\Delta t_\mathrm{A/R}$ &  $\alpha^a$ & $\Delta R$    & $r_\mathrm{max}$	\\
			 & (ms)    	     			   &  ($\degr$)	& (km)  		&	(km)			\\
	\hline
	B0329+54 & $<0.65$    			   &    30.8		& 	$<128$	&$<183^b$	\\
 	B0809+74 & $<1.28$    			   & 0.0$^c$   	& 	$<384$   	& -			\\
	B1133+16 & $<0.35$    			   &   51.3        	&  	$<59$	& $<110$		\\
	B1919+21 & $<0.28$    			   &   45.4        	& 	$<49$	&-			\\
	\hline
\end{tabular}
\tablefoot{
\tablefoottext{a}{Errors were not given for $\alpha$ in \cite{lm88}, so errors on the emission heights cannot be derived. See text for a discussion.}
\tablefoottext{b}{Emission height of inner cone.}
\tablefoottext{c}{0.0 was used to provide an upper limit because alpha is unavailable for B0809+74.}
}
\label{tab:abret}
\end{table}

Our limits agree well with previous papers, such as those by \cite{ cor78, mw80, kss11}, who also failed to detect the effects of aberration and retardation in low frequency pulsar timing data. It is also interesting to compare our findings to those of \cite{kxj+97} who performed a similar analysis at high frequencies (1.4--32~GHz) and found that  $r_\mathrm{max}< 310$~km for PSR B1133+16 and  $r_\mathrm{max}< 320$~km for PSR B0329+54. 

Our limits of $r_\mathrm{max} < 110$~km for PSR B1133+16, significantly improve upon the previous low frequency ($<200$~MHz) limits for PSR B1133+16 set by \cite{cor78} and \cite{mw80}, who found $r_\mathrm{max}< 630$~km; and \cite{kss11} who improved upon this, finding $r_\mathrm{max}< 560$~km. This is predominantly because we have a large fractional bandwidth and high sensitivity at low frequencies. Our limits are also improved by the frequency-dependent models that were used, which allowed us to test how well our fiducial points fit the data, reducing the uncertainty in $\Delta t_\mathrm{A/R}$ significantly. 

The uncertainties in our measurements are dominated by the uncertainties associated with $\alpha$, which unfortunately, are not well constrained \citep[see][for a discussion]{ew01}. No uncertainties on $\alpha$ were given in \cite{lm88}, and because of this, the uncertainties on $\Delta R$ and $r_\mathrm{max}$ are impossible to determine definitively. However, as $\alpha$ only appears in Equation~\ref{eq:abret} through a factor of $(1+\sin\alpha)$ our measurements should be within a factor of two of the true value.

The implication of these limits is that pulsar emission from all radio frequencies is produced inside a very small region of the magnetosphere\footnote{Our calculation of absolute emission height ($r_\mathrm{max}$) assumes that the emission comes from dipolar magnetic field lines emanating from the polar cap, although the calculations of the range of heights, ($\Delta R$)  is valid for any geometry.}.  All of the radio emission from B1133+16 comes from within 11 stellar radii (using the canonical neutron star radius of 10~km, as used in \citealt{kxj+97}), a fact which could have implications for future models of the pulsar magnetosphere.

\subsection{Magnetospheric Propagation Effects}
The pulsar magnetosphere is a complex system with strong magnetic fields and high concentrations of relativistic charged particles. As high frequency emission is thought to originate closer to the neutron star surface, it has more of the magnetosphere to travel through. This means that, under the assumption of radius-to-frequency mapping, we might expect to measure a slightly higher value for DM at high frequencies than at low frequencies, changing the way dispersion delay scales with frequency. 

We see no evidence for any deviation from a $\nu^{-2}$  power law in our data, suggesting that dispersion from within the magnetosphere is either not present, too small to detect, or indistinguishable from the cold plasma dispersion law (at least in the fiducial components). This suggests that either the column density of the plasma in the magnetosphere is too small to cause refraction, or emission can somehow escape the magnetosphere without being refracted. This could be because the emission of the fiducial component propagate via the extraordinary mode, which does not suffer refraction \citep{ba86}. In the pulsar magnetosphere, the electrons are very tightly bound by the magnetic field lines and cannot move transverse to their direction. This means that photons cannot excite them, effectively making them invisible and setting their refractive index to 1 \citep[see for example][]{mic91}. It should be noted that other modes of propagation, which can be refracted (ordinary modes) are also possible in the magnetosphere, although from this evidence, they do not seem to be present in the fiducial components.

\subsection{Super-dispersion}
\label{sec:superd}
Super-dispersion was proposed by \cite{sm85} after they observed that the DMs of low frequency pulsar observations seemed to be systematically higher than the DMs of the same pulsars at high frequencies. 

They explained the delay by  magnetic sweepback \citep{shi83}, in which a pulsar's magnetic field lines get bent backwards in the opposite direction to the spin of the neutron star. This causes emission from higher up in the magnetosphere (at low frequency) to reach us slightly later than the corresponding emission from closer to the neutron star surface (high frequency). This model was supported by further evidence from \cite{kuz86} who observed super-dispersion in eight more pulsars and observed a correlation between $1/P$ and the observed super-dispersive delay and also \cite{smi88} who observed the effect in pulsars B0834+06, B1133+16, B1508+55, B0832+26 and B1642$-$03, and also noted that the delay appeared to be time variable in B1133+16 and B0809+74.

In a later paper \citep{klll08} this was cast into doubt as the super-dispersion in Crab giant pulses corresponded to more than 1 period, a delay which cannot be explained by the twisting of magnetic field lines in the pulsar magnetosphere. Super-dispersion was also not seen by \cite{pw92} and \cite{agmk05}, who did, however see an excess delay at high frequencies. 

We see no evidence for extra delays at low frequencies and can place a limit on the super-dispersive delay of $\lesssim 1$~ms at 40~MHz for the pulsars in our sample. We speculate that the delay which was seen in PSR B0809+74 was actually due to either pulse profile evolution or an incorrect fiducial point (see Section~\ref{sec:prof}).

\section{Pulse Profile Evolution}
\label{sec:prof}

\subsection{Impact on Pulsar Timing}
\label{sec:timing}

We have shown that it is possible to find an analytical fiducial point for each of the pulsars in our sample, which is valid between 48~MHz and 1780~MHz, and which satisfies the $\nu^{-2}$ frequency dependence expected from the dispersive delay. Building a frequency-dependent template around this fiducial point to include the effects of pulse profile evolution across the band significantly reduces the systematic errors caused by pulse profile evolution, and improves the precision of timing observations. 

Pulse profile evolution can cause systematic errors in pulse arrival times which can be on the order of milliseconds for normal (slow) pulsars. The size of this effect is largest when the profile evolution is asymmetric, as noted in \cite{amg07}, but it still plays a role in relatively symmetric profiles like that of PSR B0329+54. In reality there are very few pulsars that have truly symmetric pulse profile evolution, so in order to obtain ever higher timing precision (on the order of microseconds for normal pulsars or tens of nanoseconds for millisecond pulsars) it is crucial to account for the evolution of pulse shape across the band. 

Although the frequency evolution across the relatively small bandwidths used up until recently will limit the effect in an individual band, it will still manifest when one tries to combine data from more than two frequencies as it cannot be absorbed into a fit for a dispersion delay.  The problem becomes more acute as we search for greater sensitivity by using wider and wider instantaneous bandwidths. Here the determination of the time of arrival itself is affected by the evolution of the profile and the dispersion delay. The method presented here provides a way in which one can use very wide band data to model the profile sufficiently to build a template which incorporates all these effects, although it remains to find the optimal way to extract a time of arrival from these data.

\subsection{Models}
To address the profile evolution induced errors in the residuals it was necessary to make frequency-dependent models for each of the pulsars in our sample. We were able to model the profile evolution of the four pulsars over seven octaves of frequency using analytic models of Gaussian fits to the data. Where necessary (in PSR B0329+54) the models were also made to include the effects of interstellar scattering. This was modelled by convolving an exponential tail (whose length was fit to the particular frequency band) with the Gaussian components. Although the models used are simple, they describe the shape of the pulse profile very well (as shown below), and are very effective in reducing the systematic errors seen in our timing residuals as a function of frequency. Our timing residuals were used to test the validity of each of our models by determining how well the frequency-dependent templates remove the different systematic errors attributed to profile evolution in each observation. The power-law-dependencies of all of the fitted parameters in the models are given in Table~\ref{tab:model_params}.

In all cases, the evolution of the relative amplitudes of the peaks was very difficult to fit. In general the peak heights within an observing band could be approximated well by a power law, but the parameters of each power law varied significantly from one observing band to the next, leading to discontinuities in our model. This could be because the pulse components do not have the same spectral index across the entire frequency range (i.e. they have one or more spectral breaks), which could cause the observed discontinuities in the `ratio of peaks' parameter. The only means to test this hypothesis requires accurate flux measurements for the various components and since absolute flux calibration is not yet available to us, the present data set cannot be used to provide a conclusive answer to this particular question.


\subsubsection{PSR B0329+54}
\begin{figure}
\centering
\includegraphics[height=0.2\textheight, trim = 2cm 1.5cm 2cm 1.5cm, clip = true]{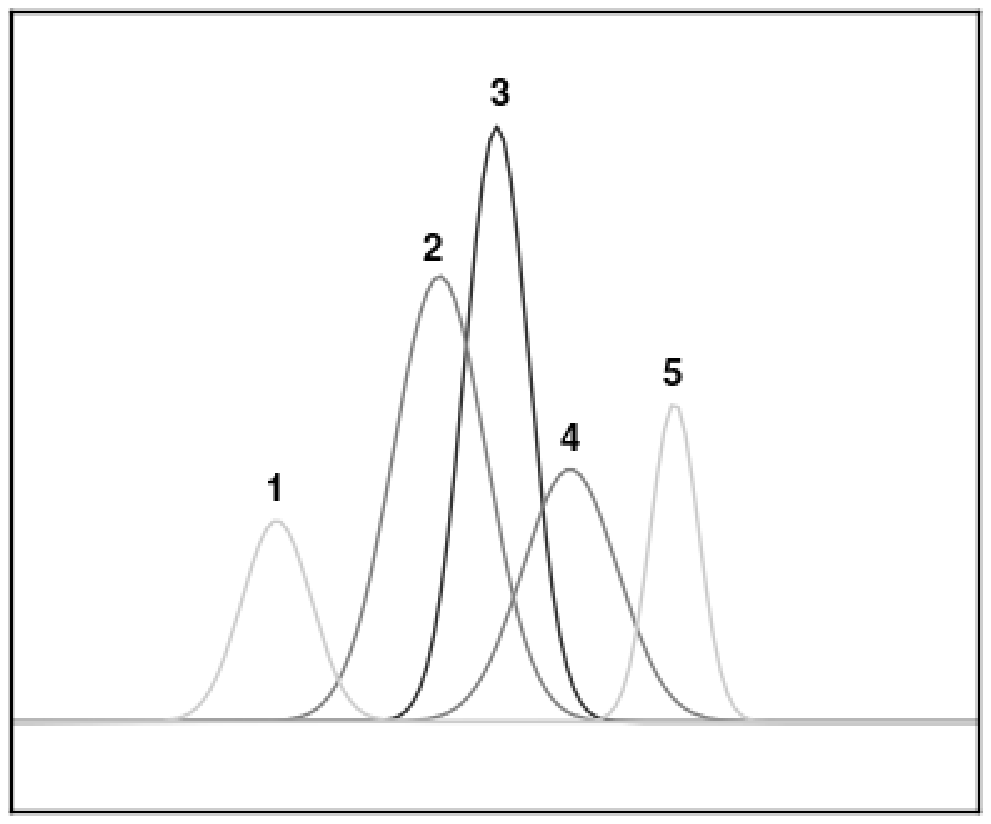}
\includegraphics[height=0.7\textheight, trim =0.8cm 0.8cm 0.8cm 2cm, clip = true]{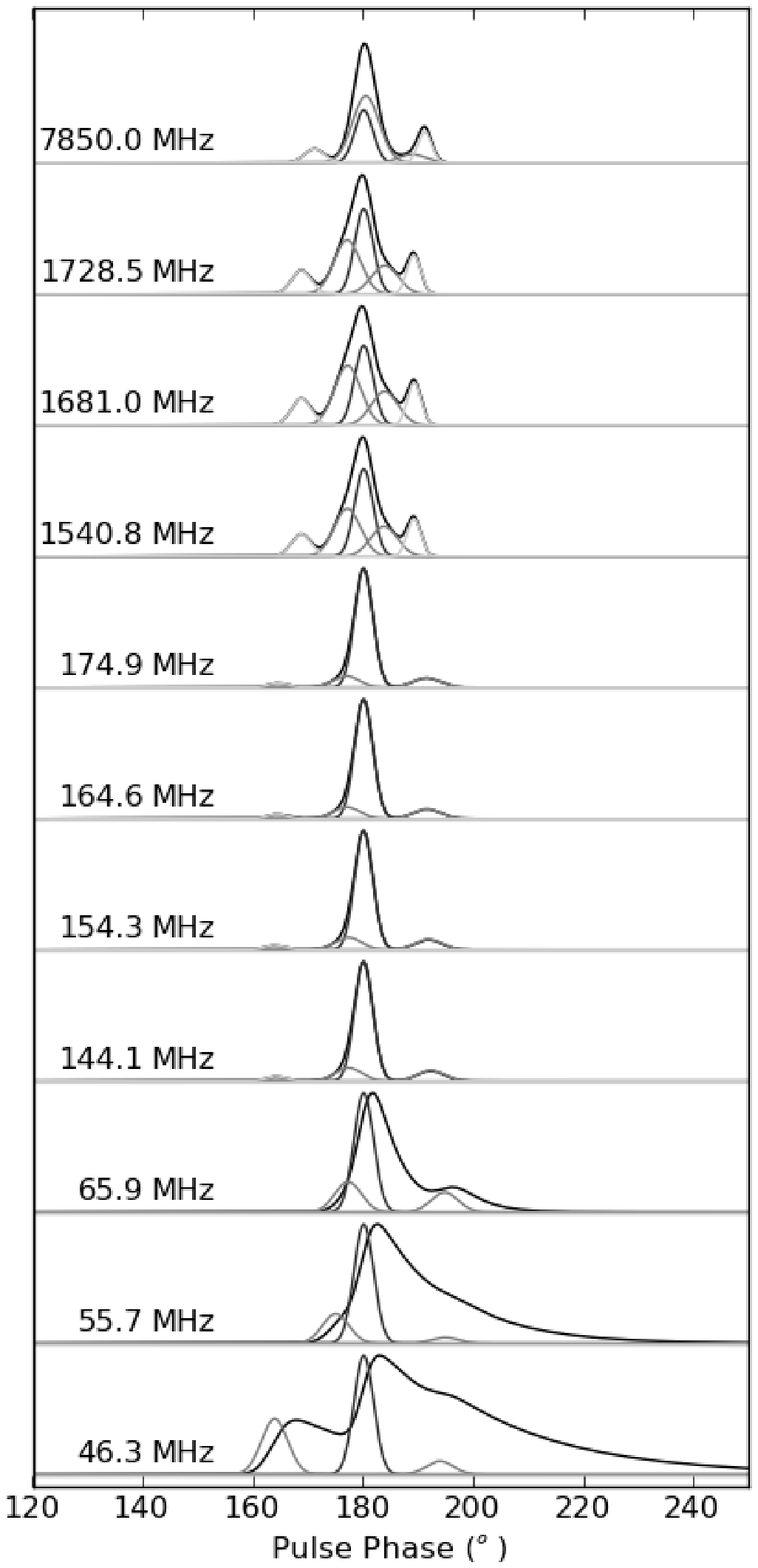}
\caption{The model of PSR B0329+54 used for the dynamic template. The pulsar is modelled using five Gaussian components (plotted with grey lines). At low frequencies the model is convolved with an exponential function to account for the effects of scatter broadening. The final model (including the scatter broadening) is plotted in black.}
\label{fig:B0329_components}
\end{figure}

\cite{gg01} (hereafter GG) model PSR B0329+54 as a central core component surrounded by four nested conal rings. Our model agrees well with this; we use a central core and two cones (five components) because rings 2 and 4 of the GG model are too weak for us to detect with any of our instruments. Figure \ref{fig:B0329_components} shows our model. 

The core component (component 3), is the fiducial point for the observation and so we have set it to be constant in height and position, allowing the other components to vary. The outer cone (components 1 and 5) fades at low frequencies, and is too weak to see in the LOFAR observations. Interestingly, the two sides of the cone fade at different rates, component 5 has a steep spectrum, and is not detected in the HBA or LBA observations (it was detected by GG at 606~MHz and 325~MHz), whilst component 1 fades more gradually, and does not disappear until the LBA observations. The inner cone (components 2 and 4) is brighter and is visible all the way down to the LBAs (although the scattering tail makes it more difficult to model the components here). Again there is evidence for the two sides of the cone showing different spectral indices. Component 2 fades more than component 4 in the high frequency observations, but component 4 fades more in the LBA band. 

The widths of all of the components seem to remain remarkably constant over the entire frequency range of our observations. In fact, PSR B0329+54 can be modelled with constant component widths from frequencies between 40~MHz and 8~GHz. It is difficult to say conclusively whether this model reflects a feature which is intrinsic to the pulsar, as the components all overlap, making them difficult to model. However a model with fixed component widths is simpler and this makes it much easier to track how the pulse profile evolves at different frequencies, how the components move and how their brightnesses vary in relation to each other. It is also worth noting that similar behaviour has been found in the Vela pulsar by \cite{kjlb11}.

\begin{figure}
\centering
\includegraphics[width=\linewidth, trim = 1cm 2cm 0cm 10cm, clip=true]{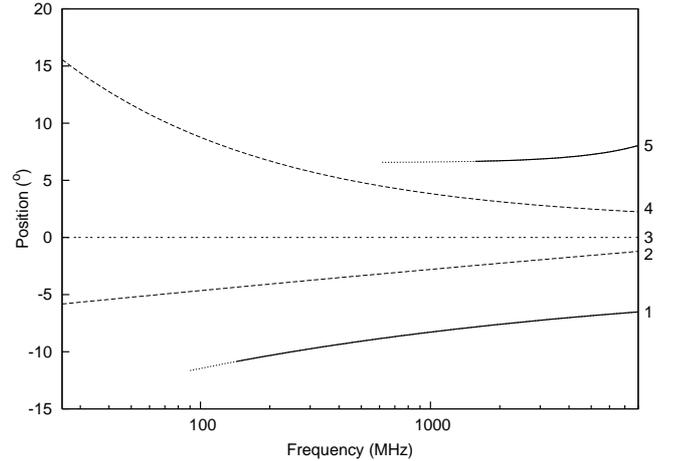}
\caption{A fit to the relative positions of the components of PSR B0329+54, which show a lot of asymmetry. The outer cone (bold line) is skewed towards earlier pulse longitudes and both of its components fade at very different rates. The inner cone (dashed line) is skewed in the opposite direction and again shows different spectral indices for each side of the cone. Models are less reliable in the LBAs ($<100$~MHz) where the pulse profile is affected by scattering.}
\label{fig:cones}
\end{figure}

Both cones are very asymmetrical in terms of the relative brightness of their two peaks and their relative positions in relation to the central component (see Figure \ref{fig:cones}). Compared to a model where the cones are symmetric around a central component, the outer cone is skewed by approximately 5~degrees towards earlier pulse longitudes. The leading component moves away from the central component with decreasing frequency, whilst the trailing component seems to move slightly closer. The inner cone is also skewed by approximately 5~degrees, but in the opposite direction. The components both move out from the main pulse,  but component 4 moves out quite quickly, and replaces component 5 as the `postcursor' in the HBAs, whilst component 2 remains in roughly the same place until the top of the LBAs, when it begins to move out. 

Our model agrees well with the GG model of a central core component surrounded by cones. The cones, however, show asymmetrical behaviour which is not expected for a dipolar magnetic field, the radio emission should come from concentric circles centred on the magnetic pole \citep[see for example][]{os76b}. 

\subsubsection{PSR B0809+74}
\begin{figure}
\centering
\includegraphics[height=0.15\textheight]{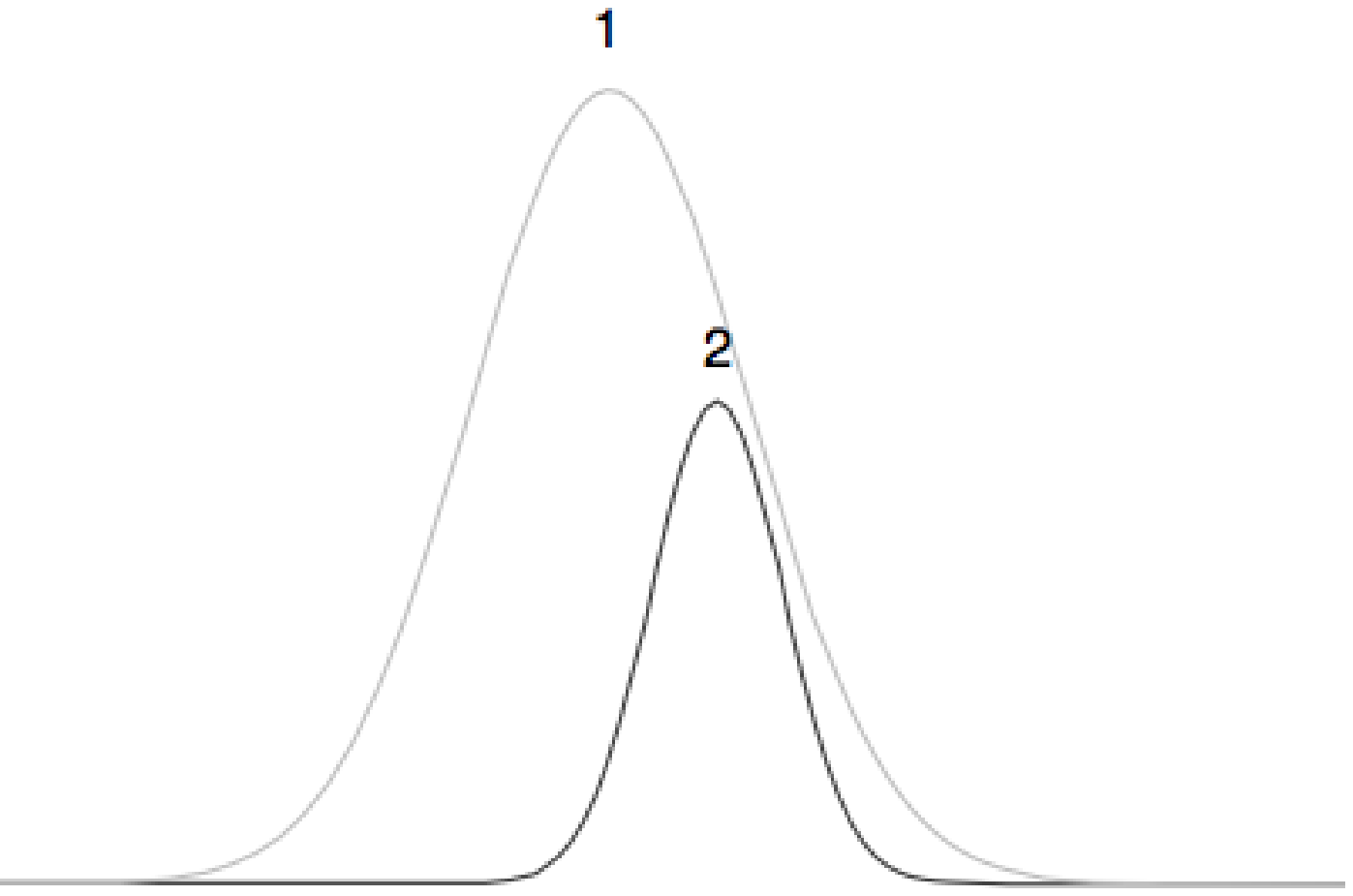}
\includegraphics[height=0.7\textheight,trim = 0cm 1.5cm 0cm 2.5cm, clip=true]{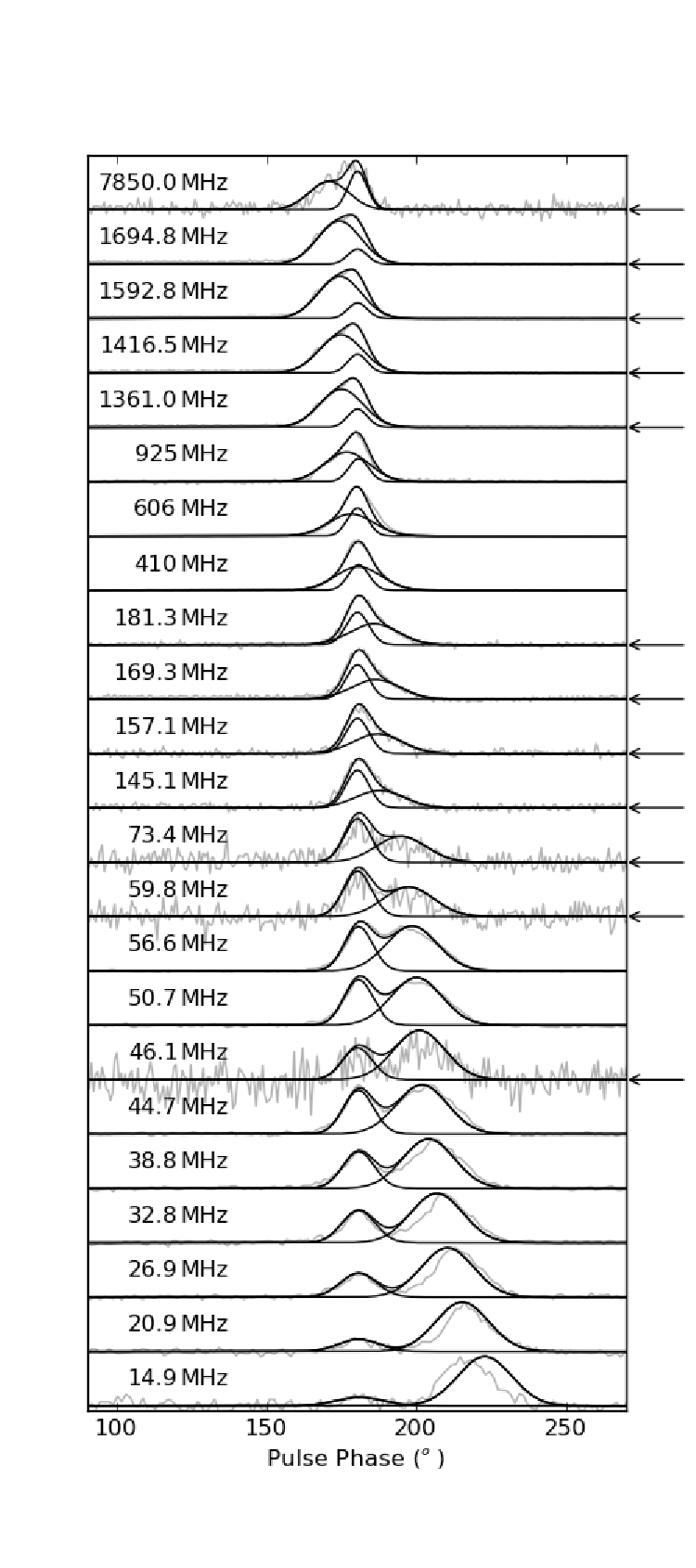}
\caption{The model used to produce the dynamic template of PSR B0809+74. The model consists of two Gaussian components. The peak of the narrower component is the fiducial point of the observation and the broad component drifts through the pulse profile. The two components and the final model are plotted in black, and compared to data, which is plotted in grey. The simultaneous observations (used to create the model) are indicated by arrows. Pulse profiles at 410~MHz, 606~MHz and 925~MHz are from the EPN database and the low frequency (10--60~MHz) pulse profiles are from a recent observation with the LOFAR superterp.}
\label{fig:B0809_components}
\end{figure}

PSR B0809+74 is a rather controversial pulsar, which has shown evidence of an `absorption feature' \citep{bar81,ran83a} and was also one of the candidates for `super-dispersion' \citep[see Section \ref{sec:superd}]{sm85}. 

Absorption was proposed by \cite{bkk+81} who noticed that the DM found in 102~MHz observations was significantly different from the value found from observations at 1720~MHz. The reason for this was that their fiducial point was on the leading edge of the low frequency pulse profile, and the trailing edge of the high frequency pulse profile. It appeared that the low frequency pulse profile was  `missing' radiation from the leading edge, which they suggested, was removed by cyclotron absorption. Further evidence for this model was also provided in a subsequent paper \citep{bar81}, which found that the profile of B0809+74 gets significantly narrower below 1~GHz. \cite{ran83a} found similar absorption features in at least eight other pulsars.

PSR B0809+74 has two overlapping components, which are normally thought to be conal. In accordance with this thinking, we fitted the data from the simultaneous observations (marked with  arrows in Figure~\ref{fig:B0809_components}) with two Gaussian components and set our fiducial point as the midpoint of the profile. This model produced large systematic errors in the TOAs at different frequencies. On a closer examination of the profiles, the reason for the timing errors became apparent, the separation of the components cannot be modelled as a simple power law. In the observations at 1400~MHz the components get closer together as frequency decreases, whereas in the low frequency data they move further apart. 

In a second model, we used three components, one central component, a precursor and a postcursor. The narrower, central component is taken as the fiducial point of the profile. At high frequencies the precursor (component 1 in Figure~\ref{fig:B0809_components}) moves towards the central component. Somewhere in the frequency range 200--1000~MHz (which was not present in the simultaneous observations) this component fades. Then, in the low frequency data, the third component appears and begins to move away from the central component, towards later pulse phase.

\begin{figure}
\centering
\includegraphics[width=\linewidth, trim=2cm 2cm -5cm 7cm, clip=true]{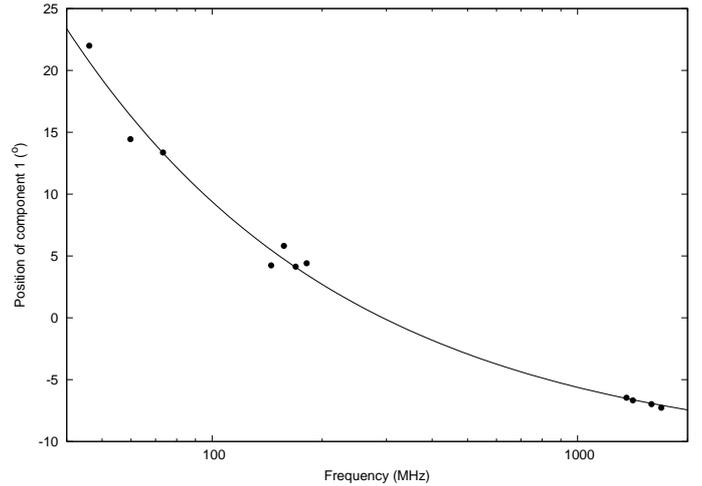}
\caption{The position of component 1 relative to component 2. The data follow a single smooth power law (which scales as $\nu^{-0.43\pm0.06}$) suggesting that the component drifts through the pulse profile.}
\label{fig:B0809_sep}
\end{figure}

The precursor and the postcursor in this model have the same width and their positions can both be modelled by a single power law (see Figure \ref{fig:B0809_sep}). This suggests that the two components may instead be a single component, which drifts through the pulse profile. 

In our final model (see Figure \ref{fig:B0809_components}), we used two Gaussian components, a narrow component (component 2), which is the fiducial point of the pulse profile, and a broader component (component 1), which starts as a precursor in the high frequency observations and drifts through the pulse profile, arriving at a later phase (as a postcursor) at low frequencies. Using the narrow component as a fiducial point removes the systematic errors from TOAs, which provides strong verification of this model. 

Further evidence in favour of this model comes from archival pulse profiles from the European Pulsar Network (EPN) database \citep{gl98}\footnote{http://www.mpifr-bonn.mpg.de/old\_mpifr/div/pulsar/data/}. These pulse profiles (at 410~MHz, 606~MHz and 925~MHz) are also shown in Figure~\ref{fig:B0809_components}, along with an interpolation of our model to these frequencies. We have allowed the relative heights of the components to vary, but their positions and widths are determined by our model. Without prior knowledge about this frequency range, our model accurately predicts the shape of the pulse profile.

The fitted solution for the profile at these frequencies (where the two components overlap), is not unique. However, it is obvious from considerations of the timing residuals that the midpoint of the two components is not the fiducial point of the pulse profile and the fact that such a simple model can explain the observed profile evolution of the pulsar over such a broad frequency range is compelling. 

We also compare our model to some more recent LBA observations taken using the LOFAR superterp\footnote{The superterp is a group of six core stations, whose signals can be combined coherently \citep[see][for more details]{sha+11}, currently the most sensitive LOFAR observing configuration for pulsars and fast transients.} (LOFAR observation ID: L30803). The data quality is significantly improved because the superterp has roughly three times the collecting area of DE601, and the delays between the dipoles have recently been calibrated. The pulse profiles are made from 6~MHz segments of bandwidth between 15~MHz and 57~MHz. Our model accurately describes the two components down to roughly 40~MHz, where the broader component begins to move away from the central component more slowly. This could be due to a mode change or some more complex pulse profile evolution at low frequencies (perhaps betraying one of the magnetospheric effects discussed above).

\begin{figure}
\centering
\includegraphics[width=\linewidth]{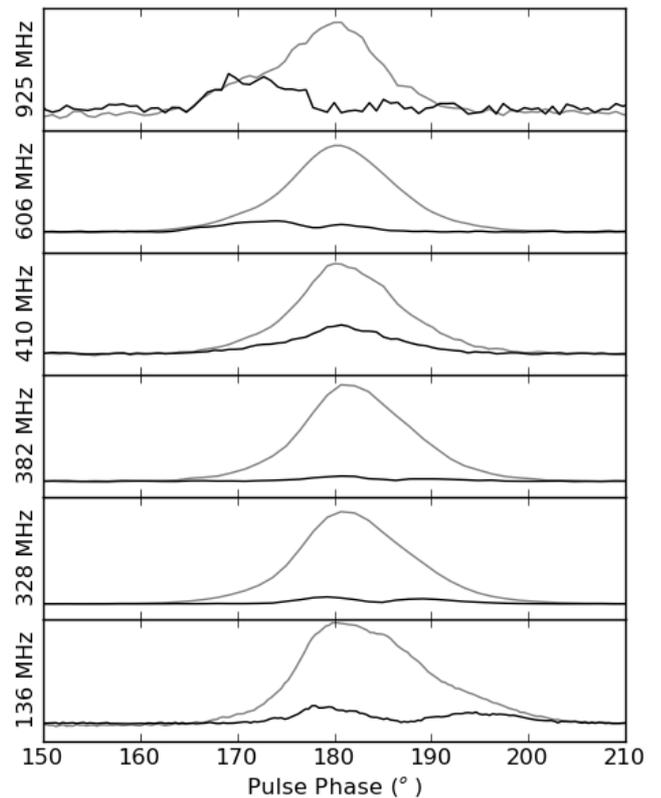}
\caption{Linear polarisation profiles of PSR B0809+74 between 925 and 328~MHz \citep{gl98,es03b}, and LOFAR polarisation profile at 136~MHz (black lines) plotted along with the Stokes I profiles at each frequency (grey lines). The polarised component moves from the position of the leading component of the profile towards later pulse longitudes, tracing the broad component of the pulse profile.}
\label{fig:epnpol}
\end{figure}

The polarisation of the EPN profiles also shows some evidence that one of the components in our model is linearly polarised (see Figure \ref{fig:epnpol}). The first component in the data is linearly polarised at high frequencies, but as frequency decreases the linear polarisation moves towards later pulse longitudes. The early LOFAR polarisation data (obs ID: L24117), shown in the bottom panel of the figure, shows that the polarisation has moved towards the latter portion of the pulse profile at 136~MHz, and arrives after the main pulse. 

One argument against this model is the step in subpulse phase which is observed at 1380~MHz   \citep[see for example][]{es03}. If the precursor at 1380~MHz is the same component that appears on the trailing edge of the profile at low frequencies, it is difficult to explain what happens to the phase jump, which is not observed at 328~MHz. The mechanism by which the two components are seen to move through each other, is also still unknown. It could be due to a retardation effect which is only present in one component, or refraction from the magnetosphere, but is not immediately obvious what could be causing the profile to evolve as it appears to. Further investigation into the single pulses of PSR B0809+74 at low frequencies could help to answer these questions.

It is interesting to note that the fiducial point in our model matches the fiducial point used by \cite{bkk+81}, which led the authors to speculate that part of the low frequency pulse profile was missing. Our model, shows that the radiation is not missing, but has been displaced somehow, to later pulse longitudes. Our model also elegantly explains the narrowing of the pulse profile (also attributed to absorption) discussed in \cite{bar81} and \cite{ran83a}. The cumulative pulse starts at high frequencies as two fairly distinct components, the components get closer together as frequency decreases, reducing the apparent width of the profile. At around 400~MHz, the profiles are exactly on top of each other, and the pulse width is at a minimum. Below this frequency, the broader pulse continues to move towards later pulse phase and the profile width begins to increase again, reproducing the shape of the absorption. 

Super-dispersion can also be explained by this model, if the centre of the two components was used as a fiducial point for the pulse profile, instead of the core component, as in our model then the pulse would look like it arrived later than expected at low frequencies. 

\subsubsection{PSR B1133+16}

\begin{figure}
\centering
\includegraphics[height=0.2\textheight, trim= 2cm 1.5cm 2cm 1.5cm, clip=true]{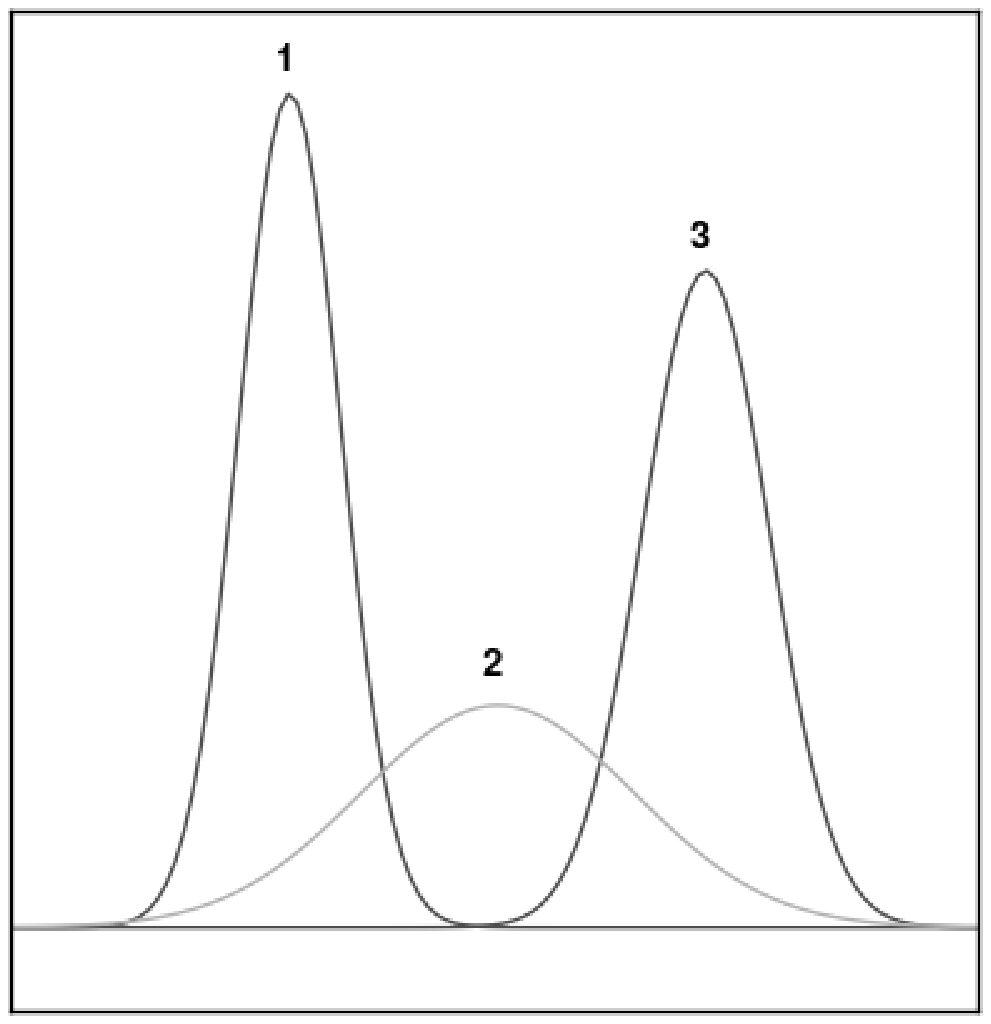}
\includegraphics[height=0.7\textheight, trim =0.8cm 0.8cm 0.8cm 2cm, clip = true ]{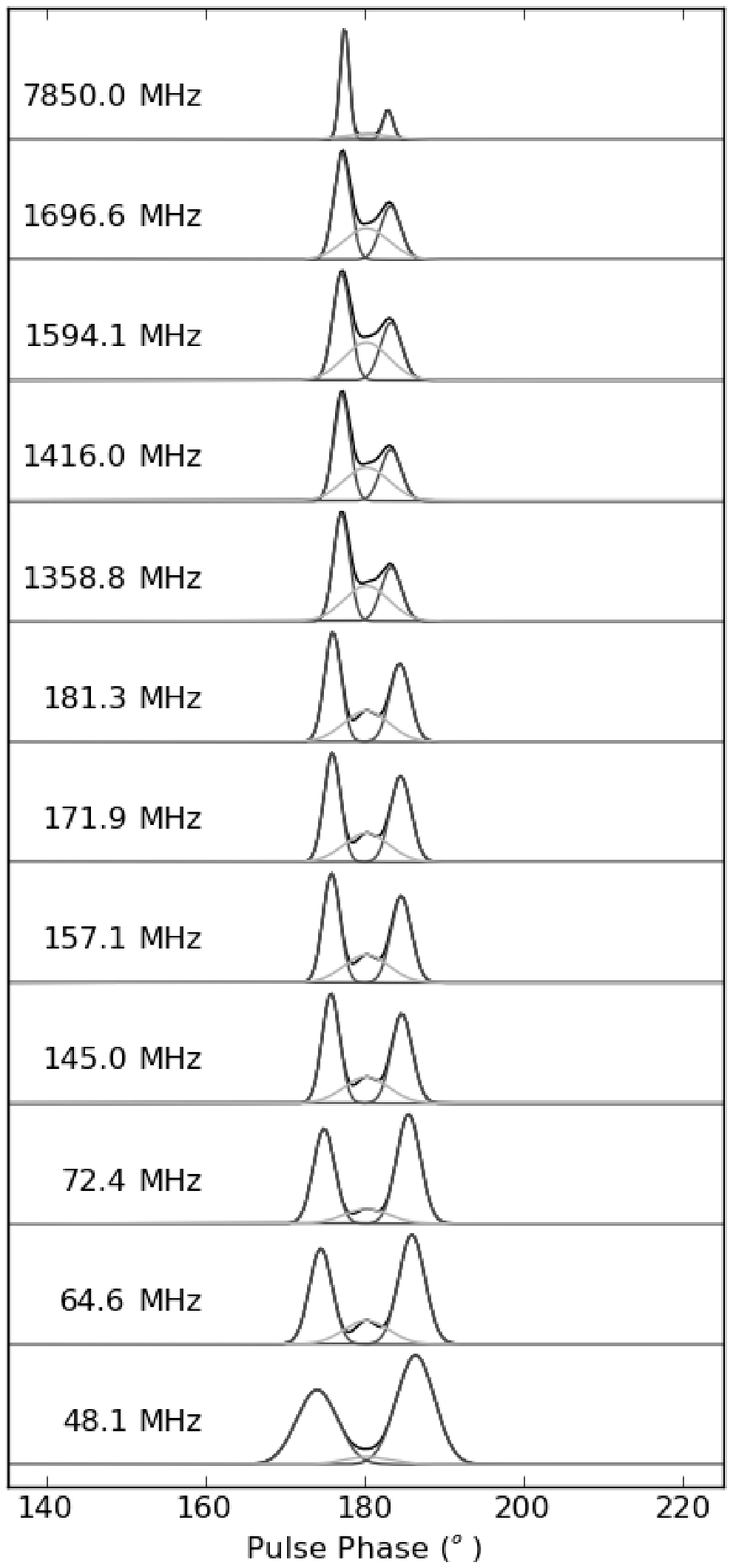}
\caption{The model used to produce the dynamic template of PSR B1133+16. The source is modelled as three Gaussian components (plotted in grey): two conal components and one which is attributed to bridge emission. The final model is also plotted in black.}
\label{fig:B1133+16_components}
\end{figure}

PSR B1133+16 is one of the prototypical examples of a `well resolved conal double' profile \citep{ran83}. It is one of the brightest pulsars in the Northern sky, so it has been widely studied and is often used as evidence in favour of radius-to-frequency mapping \citep[see for example][]{tho91a}. The separation between its components shows a continuous increase with decreasing frequency, which is thought to trace the dipolar shape of the pulsar's magnetic field. 

This is exactly what we see in our model. It has two strong components (components 1 and 3) separated by bridge emission (component 2) (see Figure \ref{fig:B1133+16_components}). The fiducial point is the midpoint between components 1 and 3, which has been defined to be the peak of the bridge emission in this model. The conal components move further apart at lower frequencies, scaling with a power law $\sim\nu^{-0.62}$.  This is consistent with the exponent found in \cite{tho91a} and \cite{xkj+96}, who found exponents of $-0.50$ and $-0.71$ respectively. The exponent is however, significantly lower than the value in \cite{kss11} and \cite{cor78}, who both found a power law  $\sim\nu^{-0.3}$. 

This is because their power law fits did not include the constant term, $\Delta \theta_\mathrm{min}$,  the width of the pulse profile at the surface of the neutron star. This term was proposed by \cite{tho91a},  who found that the separation of the components tends towards a constant value at high frequencies. If we do not include this term in our fits, we find a power law of $\nu^{-0.2}$, which is roughly consistent with those of \cite{kss11} and \cite{cor78}. For our analysis, we used the form of the power law from \cite{tho91a}, which provides a better fit to the data.

The width of the conal components as a function of frequency has been a subject of interest in the past. \cite{mr02a} found that the component widths remain constant between 40 and 3000~MHz. We also see evidence of this in our model above 80~MHz, although the component widths begin to broaden below this value. \cite{mr02a} also noticed this broadening, and attributed it to dispersive smearing across a frequency channel or scattering from the ISM. In our observations, the dispersive smearing at 48~MHz (across a single 12~kHz channel) is $\sim1.5^\mathrm{o}$, which is enough to explain the observed broadening in the profile\footnote{The half power width of component~1 increases from 1.9$^\mathrm{o}$ at 72~MHz to 3.6$^\mathrm{o}$ at 48~MHz, and component~2 increases from 2.1$^\mathrm{o}$ at 72~MHz to 3.4$^\mathrm{o}$ at 48~MHz.}. However, even disregarding this low frequency broadening \cite{mr02a} found that the spacing between the components between 100~MHz and 10~GHz changes too rapidly to be caused by a dipolar magnetic field. 

PSR B1133+16 is the pulsar which is most consistent with radius-to-frequency mapping of all the pulsars in our sample. However, in Section \ref{sec:mag} we showed that its emission is confined to a very narrow region in the magnetosphere ($<$59~km) which is incompatible with the standard radius-to-frequency model. Radius-to-frequency mapping assumes that the emission at a given emission height traces the last open field line in the pulsar magnetosphere. From geometrical arguments \citep[see for example][]{lk05} it is possible to write the opening angle of the last open field line of a non-relativistic dipole, $\rho$, as a function of emission height, $r$:
\begin{equation}
\rho = 86^\circ \frac{r}{R_\mathrm{LC}},
\end{equation}
where $R_\mathrm{LC}$ is the radius of the light cylinder of the pulsar. $\rho$ is the maximum value that it is possible for components to be separated by as each component originates from one side of the dipolar field. The maximum increase in component separation predicted by this equation for PSR B1133+16 over the height range, $\Delta R$, derived earlier is 2.8$^\circ$.  In the pulse profile of PSR B1133+16 the components are seen to move apart by 6.3$\pm0.5^\circ$ between 1780~MHz and 48~MHz. This, coupled with the further evidence by \cite{mr02a} suggests that PSR B1133+16 cannot be explained through radius-to-frequency mapping in a simple dipolar magnetic field.

\subsubsection{PSR B1919+21}
\label{sec:B1919+21}
PSR B1919+21 was the first pulsar ever discovered \citep{hbp+68} and is often referred to as a classic radio pulsar. In fact, PSR B1919+21 disagrees with the classic picture of a pulsar in almost all aspects of its pulse profile evolution.

\cite{lsg71} showed that the two components of the pulsar get closer together as frequency decreases in the range 150--3000~MHz, which is the opposite to what one would expect from radius-to-frequency mapping. This result was confirmed by \cite{srw75} who also showed that the pulse profile seems to get broader again below 150~MHz. \cite{mr02a} found that the width of the profile and separation of the components was approximately constant in their observations between 50 and 5000~MHz. They suggested that the lack of any radius-to-frequency mapping could be due to the emission originating from an `inner cone', which is located closer to the `core' emission component, and so the effects of radius-to-frequency mapping are less pronounced.

In our model (see Figure \ref{fig:Fitting}) the two components move closer together at low frequencies \citep[in agreement with][]{lsg71}. We did not, however, see the component separation increasing below 150~MHz \citep[as reported by][]{srw75}.  This behaviour is similar to that of PSR B0809+74 at high frequencies, although in this case the components are only expected to pass through each other at $\sim1$~MHz. Another curious feature of our model is that whilst component 2 gets broader at low frequencies (as expected), component 1 appears to get narrower. Radius-to-frequency mapping suggests that both components should get broader at low frequencies, as the emission region gets wider. In an attempt to conform with this idea, we tried to fit a three-component model. This fit the pulse profile reasonably well, and also agreed better with the radius-to-frequency model;  the components didn't move closer together, and stayed at a constant width. However, a suitable fiducial point could not be determined, and there were large systematic errors in our timing residuals so the model was rejected.

\begin{table}
\caption{Parameters of the models used in the dynamic templates. The functions given are appropriate for frequencies measured in MHz, and pulse phase measured in degrees. Component brightnesses are defined relative to a fiducial component.}
\begin{center}
\begin{tabular}{l|l|l}
\hline
{\bf PSR B0329+54} & & \\
\hline
Component 1 	& Position 	&	$-\nu^{-0.22}$ \\
			& Amplitude	&	$\nu^{-0.018}$ \\
			& Width		&	1.7\degr \\
Component 2 	& Position 	&	$\nu^{-0.022}$\\
			& Amplitude	&	$\nu^{0.48}$\\
			& Width		&	2.4\degr \\
Component 3$^a$ 	& Position 	& $\nu^{0}$ \\
			& Amplitude	&	$\nu^{0}$ \\
			& Width		&	1.8\degr \\
Component 4 	& Position 	&	$-\nu^{0.015}$\\
			& Amplitude	&	$\nu^{0.022}$\\
			& Width		&	2.4\degr \\
Component 5	& Position 	&	$\nu^{1.5}$\\
			& Amplitude	&	$\nu^{0.48}$\\
			& Width		&	1.2\degr \\
\hline
{\bf PSR B0809+74} & & \\
\hline
Component 1 	& Position		& 	$\nu^{-0.43}$ \\
			& Amplitude ($<100$~MHz)	& 	$-\nu^{-0.055}$ \\
			& Amplitude ($>100$~MHz)	& 	$-\nu^{0.013}$\\	
			& Width		&	14.9\degr \\
Component 2$^a$& Position	& 	$\nu^{0}$	\\
			& Amplitude 	& 	$\nu^{0}$ 	\\
			& Width		&	7.7\degr \\			
\hline
{\bf PSR B1133+16}& & \\
\hline
Component 1 	& Position		& 	$\nu^{-0.62}$\\
			& Amplitude	& 	$\nu^{-3.28}$\\
			& Width		&	$\nu^{-3.55}$\\
Component 2$^a$& Position	&	 $\nu^{0}$ \\
			& Amplitude	& 	 $\nu^{0}$ \\
			& Width		&	3.2\degr \\	
Component 3   & Position		&	$\nu^{-0.62}$ \\
			& Amplitude	& 	$\nu^{-3.11}$\\
			& Width		&	$\nu^{-3.98}$\\			
\hline
{\bf PSR B1919+21}$^b$& & \\
\hline
Component 1 	& Width			& $\nu^{0.04}$	\\
Component 2	& Width			& $-\nu^{0.01}$	\\			
Separation 	& 				& $\nu^{0.05}$	\\
Ratio of Peaks	& $\nu <100$~MHz					& $\nu^{-0.07}$	\\
			& $100\mathrm{~MHz}<\nu<300$~MHz	& $\nu^{0.54}$	\\
			& $\nu > 300$~MHz					& 1.2 	\\
\hline
\end{tabular}
\end{center}
\label{tab:model_params}
\tablefoot{
\tablefoottext{a}{Component used as the fiducial component.}
\tablefoottext{b}{Midpoint between components used as fiducial point.}
}
\end{table}%

\section{Discussion}
\label{sec:disc}
\subsection{Profile Evolution}

We have shown that pulse profile evolution can introduce large errors into pulsar timing data, in agreement with the work of \cite{amg07}. These errors can be as large as a few milliseconds in some cases, depending on the period of the pulsar, and how asymmetric the pulse profile is. A frequency-dependent model of the pulse profile can be used to reduce these errors. Using this method, it was possible to define an analytical fiducial point in the pulse profile of each of the pulsars in our sample. This fiducial point is valid to within a few milliseconds (corresponding to $\sim1$ degree in pulse phase), although the model does not remove the timing errors completely. Small timing errors remain because the model is not an exact fit to the observed profile, and subtle differences between the shape of the modelled templates and the data lead to systematics in the cross-correlation. Further work needs to be done to explore how to better remove these timing errors, and how they vary with time. 

We have found that radio emission from all of the pulsars in our sample originates from a narrow range of heights in the magnetosphere. The narrow ranges found do not fit well with models of radius-to-frequency mapping in a dipolar magnetic field. In addition, all of the pulsars in our sample show at least one other trait which cannot be explained by radius-to-frequency mapping.

The asymmetric cones which we observe in PSR B0329+54 were also observed by GG, who attributed their asymmetry to aberration and retardation. However, we have not detected any aberration or retardation effects in our timing residuals and we also find that the emission from the inner cone seems to be concentrated to within 183~km of the neutron star surface. The ($\sim 5$~degree) skew in the cone corresponds to a time difference of $\sim 10$~ms, which is much greater than the aberration and retardation effects possible from within this height range, which are $\sim 1$~ms (from Equation~\ref{eq:abret}). The fact that the outer cone is skewed in the opposite direction to the inner cone, also suggests that this cannot be explained by the standard model of a pulsar.

PSR B0809+74 has a component which starts out as a precursor at high frequencies and then drifts through the centre of the pulse profile, swapping sides with the central component and appearing as a postcursor at low frequencies. The frequency dependence of the position of the drifting component suggests that either refraction or some relationship between frequency and height (a change in height could explain a component being delayed) significantly influences one component, but is not seen in the other.

PSR B1133+16 shows emission from a very narrow range of heights and, as \cite{mr02a} showed, component separation which increases too rapidly to be produced by dipolar field lines. One explanation for this could be that there are other mechanisms at work, which act together with the traditional picture of a pulsar, complicating pulse profile evolution. 

PSR B1919+21 has a profile whose width decreases at lower frequencies. This is the exact opposite of what is predicted by radius-to-frequency mapping, and so is very difficult to explain using the standard picture of a pulsar. Again, there is a clear relationship between pulse shape and frequency, but it does not seem to be explainable by radius-to-frequency mapping.

The fact that none of the pulsars in our sample behave as predicted by radius-to-frequency mapping suggests that a more complicated model of the pulsar magnetosphere is needed to describe pulse profile evolution. Although radius-to-frequency mapping has been successful in explaining some of the features seen in pulse profiles, it is clear that it cannot be used to fully describe any of the pulsars in our sample. There are, however, alternative theories which could potentially provide good fits to observational data.

In the family of models proposed by \cite{ba86}, and developed further by \cite{pet00}, \cite{wsve03} and \cite{bp11}, the frequency-dependent profile evolution seen in pulsars is explained by propagation effects in the pulsar magnetosphere. In these models, the radiation originates from a small region in the magnetosphere, and refraction, dispersion and the different propagation modes (i.e. extraordinary and ordinary) in the magnetosphere are responsible for the frequency evolution of the different components which are observed in pulse profiles. 

\cite{kj07} also provide an interesting empirical model of the pulsar magnetosphere, which could be used to explain all of the features which we have observed. They postulate that all radio emission originates from a patchy cone bounded by the last open field lines and that emission can come from any height, independently of frequency. Complex pulse profiles can then be explained by invoking emission from a range of heights, rather than assuming that the pulse profile probes the longitudinal shape of the beam at one single height.

What causes pulse profile evolution is still an important question, and will be vital in understanding the pulsar emission mechanism, and for studies of pulsar geometries in the future. At this stage, it is still difficult to discriminate between the many models that exist, but next generation telescopes, like LOFAR, will be excellent tools for studying this effect.

\subsection{Magnetospheric Effects}
The argument that a more sophisticated model is needed to describe radio emission from pulsars is also supported by considerations of aberration and retardation effects on our data. From these arguments, we have shown that radio emission from all of the pulsars in our sample is confined to a very small region in the pulsar magnetosphere, which supports the ideas of \cite{ba86} and \cite{pet00}. 
 
We have also shown that, as there is no departure from a $\nu^{-2}$ dispersion law in our data, there is no evidence for super-dispersive delays or refraction from within the pulsar magnetosphere in any of the pulsars in our sample. However, whilst we don't see a frequency-dependent delay in the timing residuals, refraction may be needed to explain the broad component of PSR B0809+74, which is seen to drift through the pulse profile. 

\subsection{ISM Effects}
In our data, we see no evidence for any deviation from the cold plasma dispersion law, suggesting that  second-order ISM effects in these pulsars introduce additional time delays $\lesssim50$~ns at normal pulsar timing frequencies (1400~MHz). The parameters of the pulsars in our sample are typical of those found in the PTAs, the only difference being their longer pulse periods. This suggests that the ISM may not cause as much of a hindrance to pulsar timing projects as first feared \citep{hs08}.

The fact that no unexpected delay was detected in any of our observations also means that (at least along these lines of sight) the ISM appears to be relatively smooth, with no large, dense structures. These findings agree with the idea that scattering is dominated by one or two small, but relatively high density regions as discussed by \cite{sti06} and \cite{bmg+10}. We have determined an upper limit on $\frac{d}{dx} \mathrm{DM}$, which can be used as an indicator as to how much the DM is likely to change in the future. Comparing these predictions with reality will be a useful cross-check of how well this relation works. We have also been able to place upper limits on the scattering measure and the emission measure. Although these limits are weak compared to NE2001 \citep{cl02}, pulsar timing could provide an independent method of measuring both of these quantities in the future. 

\subsection{Future Observations}
The constraints set in this paper will be improved significantly by taking similar observations when LOFAR is completed, using more stations at a lower observing frequency. The sensitivity of LOFAR has already improved by a factor of five since the observations for this paper were taken, and it is expected that it will increase significantly again soon, when the core stations can all be combined coherently. Increased sensitivity, particularly in the LBAs, will reduce the error bars seen at low frequencies in our timing residuals, which dominate the uncertainty in our measurement. 

We could also increase our precision by observing at lower frequencies. LOFAR will soon be able to routinely observe with high sensitivity at frequencies as low as 15~MHz (see Figure~\ref{fig:B0809_components}), where the second-order ISM delays are expected to be at least an order of magnitude larger than they are at 40~MHz. 

By repeating this experiment in the future on the same set of pulsars, we could test whether pulse profile evolution is stable with time, and also track variations in the DM with great accuracy. Both of these parameters are not completely understood, and are vital for high-precision pulsar timing. It would also be interesting to perform this experiment on a millisecond pulsar. A faster rotation rate and a narrower pulse (in absolute terms) means that TOAs can be determined more accurately, which would improve our constraints by at least an order of magnitude. 

\section*{Acknowledgements}
We would like to thank Jim Cordes for his insight and useful discussions, Christine Jordan for arranging the observations from Jodrell Bank, and the anonymous referee for their insightful comments. LOFAR, the LOw Frequency ARray designed and constructed by ASTRON, has facilities in several countries, that are owned by various parties (each with their own funding sources), and that are collectively operated by the International LOFAR Telescope (ILT) foundation under a joint scientific policy. This publication made use of observations taken with the 100-m telescopes of the MPIfR (Max-Planck-Institut f\"ur Radioastronomie) at Effelsberg. Ben Stappers, Patrick Weltevrede and the Lovell observations are supported through an STFC rolling grant. 
Tom Hassall is the recipient of an STFC studentship. Jason Hessels is a Veni Fellow of the Netherlands Foundation for Scientific Research. Joeri van Leeuwen and Thijs Coenen are supported by the Netherlands Research School for Astronomy (Grant NOVA3-NW3-2.3.1) and by the European Commission (Grant FP7-PEOPLE-2007-4-3-IRG \#224838). Aris Karastergiou is grateful to the Leverhulme Trust for financial support. Joris Verbiest is supported by the European Union under Marie Curie Intra-European Fellowship 236394. Charlotte Sobey is supported by the DFG (German Research Foundation) within the framework of the Research Unit FOR 1254,
ÒMagnetisation of Interstellar and Intergalactic Media: The Prospects of Low-Frequency Radio ObservationsÓ.


\bibliography{psrrefs}
\bibliographystyle{aa}
\end{document}